\documentclass[twoside,twocolumn,9pt]{article}

\usepackage{extsizes}
\usepackage[super,sort&compress,comma]{natbib} 
\usepackage[version=3]{mhchem}
\usepackage[left=1.5cm, right=1.5cm, top=1.785cm, bottom=2.0cm]{geometry}
\usepackage{balance}
\usepackage{mathptmx}
\usepackage{sectsty}
\usepackage{graphicx} 
\usepackage{lastpage}
\usepackage[format=plain,justification=justified,singlelinecheck=false,font={stretch=1.125,small,sf},labelfont=bf,labelsep=space]{caption}
\usepackage{float}
\usepackage{fancyhdr}
\usepackage{fnpos}
\usepackage[english]{babel}
\addto{\captionsenglish}{%
  
}
\usepackage{array}
\usepackage{droidsans}
\usepackage{charter}
\usepackage[T1]{fontenc}
\usepackage[usenames,dvipsnames]{xcolor}
\usepackage{setspace}
\usepackage[compact]{titlesec}
\usepackage{hyperref}
\usepackage{siunitx}
\usepackage{mhchem}
\DeclareSIUnit{\molar}{\textsc{m}}
\usepackage{epstopdf}
\usepackage{newfloat}
\DeclareFloatingEnvironment[name={Supplementary Figure}]{suppfigure}
\definecolor{cream}{RGB}{222,217,201}

\begin{document}

\makeFNbottom
\makeatletter
\renewcommand\LARGE{\@setfontsize\LARGE{15pt}{17}}
\renewcommand\Large{\@setfontsize\Large{12pt}{14}}
\renewcommand\large{\@setfontsize\large{10pt}{12}}
\renewcommand\footnotesize{\@setfontsize\footnotesize{7pt}{10}}
\makeatother
\newcommand \zz {\textcolor{blue}}
\newcommand \jpb {\textcolor{magenta}}
\renewcommand \fs {\textcolor{red}} 
\renewcommand{\thefootnote}{\fnsymbol{footnote}}
\renewcommand\footnoterule{\vspace*{1pt}%
\color{cream}\hrule width 3.5in height 0.4pt \color{black}\vspace*{5pt}} 
\setcounter{secnumdepth}{5}

\makeatletter 
\renewcommand\@biblabel[1]{#1}            
\renewcommand\@makefntext[1]%
{\noindent\makebox[0pt][r]{\@thefnmark\,}#1}
\makeatother 
\renewcommand{\figurename}{\small{Fig.}~}
\sectionfont{\sffamily\Large}
\subsectionfont{\normalsize}
\subsubsectionfont{\bf}
\setstretch{1.125} 
\setlength{\skip\footins}{0.8cm}
\setlength{\footnotesep}{0.25cm}
\setlength{\jot}{10pt}
\titlespacing*{\section}{0pt}{4pt}{4pt}
\titlespacing*{\subsection}{0pt}{15pt}{1pt}

\makeatletter 
\newlength{\figrulesep} 
\setlength{\figrulesep}{0.5\textfloatsep} 

\newcommand{\topfigrule}{\vspace*{-1pt}%
\noindent{\color{cream}\rule[-\figrulesep]{\columnwidth}{1.5pt}} }

\newcommand{\botfigrule}{\vspace*{-2pt}%
\noindent{\color{cream}\rule[\figrulesep]{\columnwidth}{1.5pt}} }

\newcommand{\dblfigrule}{\vspace*{-1pt}%
\noindent{\color{cream}\rule[-\figrulesep]{\textwidth}{1.5pt}} }

\makeatother

\twocolumn[
  \begin{@twocolumnfalse}
{}\par
\vspace{1em}
\sffamily
\begin{tabular}{m{1.5cm} p{13.5cm} }

& \noindent\LARGE{\textbf{Light-activated microtubule-based 2D active nematic}} \\
\vspace{0.3cm} & \vspace{0.3cm} \\

 & \noindent\large{Zahra Zarei,\textit{$^{a}$} John Berezney,\textit{$^{a}$} Alexander Hensley,\textit{$^{a}$} Linnea Lemma,\textit{$^{b}$} Nesrin Senbil,\textit{$^{a}$} Zvonimir Dogic,\textit{$^{c}$} and Seth Fraden,\textit{$^{a}$}} \\ \\

& \noindent\normalsize{ We characterize two-dimensional (2D) microtubule-based active nematics driven by light-responsive kinesin motor clusters. We assess two constructs of optogenetic kinesin: opto-K401, a processive motor, and opto-K365, a non-processive motor. Measurements reveal an order of magnitude improvement in the contrast of nematic flow speeds between maximally- and minimally-illuminated states for opto-K365 motors. Focusing on opto-K365 nematics, we characterize both the steady-state flow and defect density as a function of applied light and examine the transient behavior between steady-states. The steady-state nematic flow and defect densities are set by the applied light intensity across centimeter-sized samples, independent of initial conditions. Although nematic flow reaches steady-state within tens of seconds, the defect density exhibits transient behavior for $4$ to $10$ minutes, showing a separation between small-scale active reorganization and system-scale structural states. This work establishes an experimental platform to test theoretical frameworks which exploit spatiotemporally-heterogeneous patterns of activity to generate targeted dynamical states. }
\end{tabular}
\vspace{0.6cm}
\end{@twocolumnfalse}] \vspace{0.6cm}


\renewcommand*\rmdefault{bch}\normalfont\upshape
\rmfamily
\section*{}
\vspace{-1cm}


\footnotetext{\textit{$^{a}$~The Martin Fisher School of Physics, Brandeis University, Waltham, Massachusetts 02454, USA. E-mail: fraden@brandeis.edu}}

\footnotetext{\textit{$^{b}$~The Department of Chemical and Biological Engineering, Princeton, NJ 08544, USA. }}

\footnotetext{\textit{$^{c}$~Department of Physics, University of California, Santa Barbara, California 93106, USA. }}


\section{Introduction}
Active matter nematics are liquid crystalline materials that use an internal energy source to power autonomous chaotic flows. They are a paradigmatic example of an internally driven non-equilibrium system and have diverse potential applications ranging from microfluidics to autonomous soft robotics. Unbounded active nematics are intrinsically unstable, which limits the potential for applications~\cite{simha2002hydrodynamic,Sanchez2012,decamp,zhou2014living}. Theory predicts that confinement of 2D active nematics by hard boundaries and by gradients in viscous dissipation to the third dimension can produce robust flows with a variety of spatiotemporal patterns. Experiments support the notion that boundaries produce ordered flows, although it should be noted that no amount of confinement has fully suppressed the uncontrolled creation of topological defects\cite{Deforet2014,Keber2014,Opathalage20194788,Thijssen2021,Zhao2020,Hardoüin2019,Chandrakar2020,Rivas20209331}. While boundaries can structure active flows, applications additionally require a mechanism to switch flows between different states.  Recent advances have demonstrated the possibility of using external optical signals to control the spatial activity in 3D contractile and extensile active fluids as well as 2D actin-based active nematics \cite{ross2019controlling,zhang2021spatiotemporal,lemma2022spatiotemporal}. 

Light-activated nematics pave the way for new research directions. For example, in a recent theoretical study, optimal control theory was employed to compute the spatiotemporal patterning of activity necessary to switch confined nematics from one stable state to another\cite{Norton2020}. Other work has focused on optimal designs for moving droplets of active nematics through space and creating specific patterns of defects within a material \cite{shankar2022optimal,shankar2022spatiotemporal,falk2021learning}. More ambitious control goals are desirable. For instance, stabilizing the active nematic bend instability at a point partially through the transition from aligned to turbulent, or reversing the turbulent state to an ordered one. Here, we assess the capability of an active 2D nematic of extensile microtubule bundles with light-sensitive kinesin motors to experimentally test optimal and other forms of control theory. We focus on two properties of the light-sensitive active nematics; (1) the range of speeds of the active nematic controlled by light and the transient response time, and (2) the range of defect densities controlled by light and the transient response time. 

\section{Light-sensitive MT-based active nematics}
Conventional microtubule (MT)-based active fluids consist of orientationally aligned micron-long microtubule filaments that lack large-scale polar order\cite{Sanchez2012, chandrakar2022engineering, Lemma2019}. Clusters of truncated kinesin-1 molecular motors can simultaneously bind multiple filaments and can generate relative sliding for anti-parallel microtubules. If the cluster is not sufficiently rigid, or if the motors become unlinked, then each motor walks independently and no stress is developed. Only when the motors are rigidly linked can  the  clusters induce interfilament sliding, which leads to the filamentous bundle extending parallel to the filament axis and thinning perpendicular to the axis\cite{lemma2020multiscale}. Such mesoscale motion  scale of MT filaments is transferred to the macroscopic scale and the active nematic is observed to elongate parallel to the nematic director and contract in the perpendicular direction\cite{lemma2020multiscale}. 

\begin{figure}[h]
\centering
\includegraphics[height=5cm]{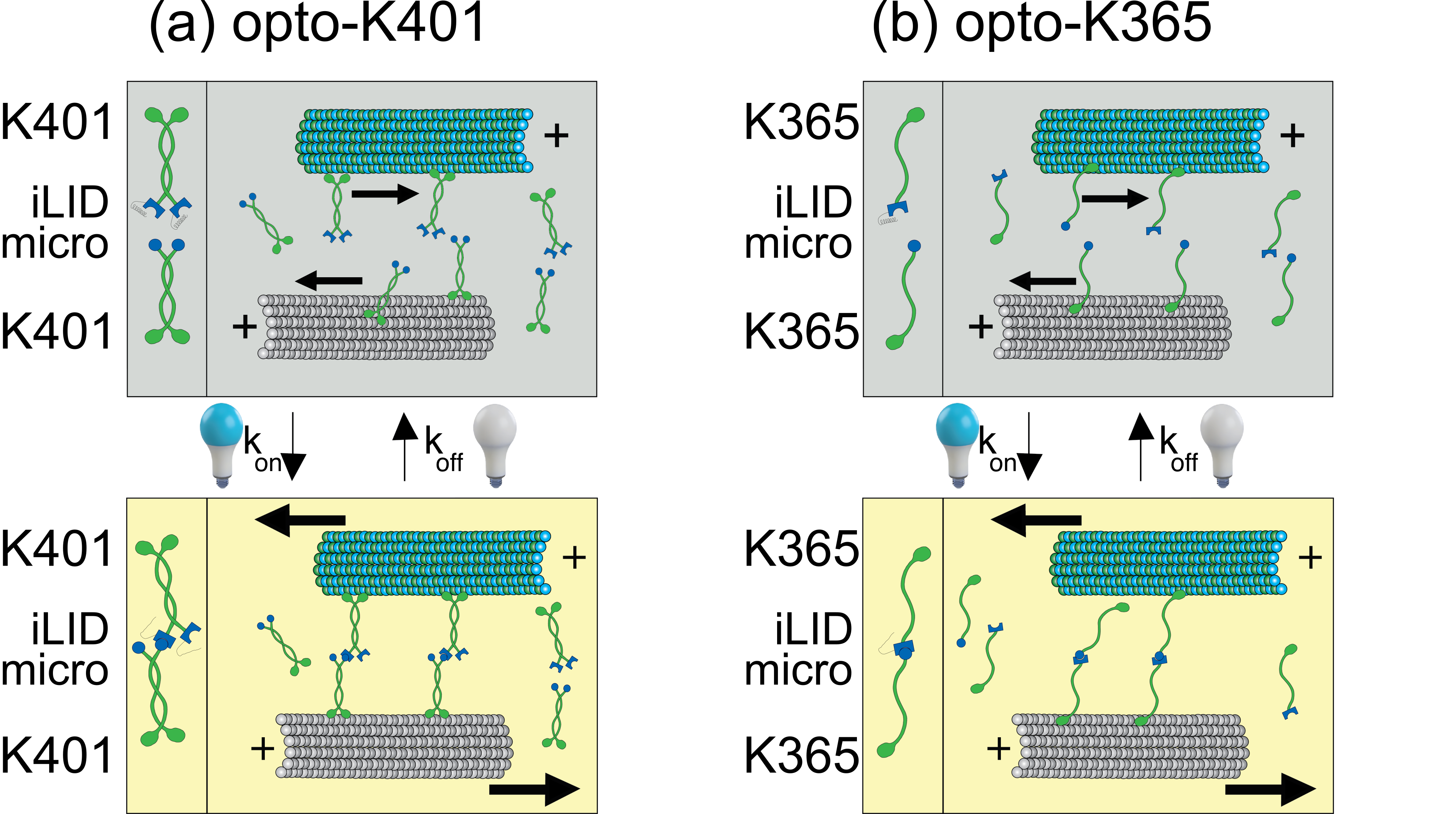}
  \caption{The optogenetic domains, iLID and micro, are fused to kinesin (K) motors K401 and K365. \textit{Top row}: In the dark state, K-iLID and K-micro are unlinked. The motors move towards the microtubule plus-end and the microtubules are stationary. (\textbf{a})  K401-iLID and K401-micro form processive homodimers.   (\textbf{b}) K365-iLID and K365-micro are monomeric and non-processive. \textit{Bottom row}: When the proteins are illuminated with blue light, the iLID domain changes conformation, enabling formation of a kinesin hetero-dimer capable of binding two microtubules. The motor clusters are stationary and the microtubule plus-ends move towards the motor cluster, generating interfilament sliding. } 
  \label{fgr:LemmaOptoKinesinSchematic}
\end{figure}

Light-sensitive kinesin-1 rigid clusters designed to power extensile active microtubule bundles have recently been developed and used in studies of active isotropic gels (Fig. \ref{fgr:LemmaOptoKinesinSchematic})~\cite{Tyler2019,LemmaL_arXiv_2022}. We use the same clusters to create light-responsive 2D MT-based active nematics. Kinesin motors are imbued with light sensitivity through fusion with optically responsive hetero-dimerizing domains, one of which is called ``iLID" (improved light-induced dimers), while the other is called ``micro''. In the absence of blue light, the binding affinity of iLID to its polypeptide binding partner, micro, is low. In the presence of blue light, iLID and micro bind rigidly to each other with a 50-fold increase in affinity\cite{iLID2015}. To create a light-sensitive cluster, two kinesin chimeras were formed; one kinesin fused with iLID (K-iLID) and one fused with micro (K-micro). Rigid kinesin clusters form when these species are mixed together in the presence of blue light.

We used two designs of iLID-micro kinesin clusters (Fig. \ref{fgr:LemmaOptoKinesinSchematic}). The first, opto-K401, is based on the 401 amino-acid fragments of Drosophila kinesin-1 motor (K401)\cite{Nedelec1997,Sanchez2012}. Two K401 proteins associate to form a two-headed homo-dimeric kinesin motor. These homo-dimeric motors are processive, remaining continuously bound to a single microtubule as they take about one hundred 8-nm “hand-over-hand” steps toward the microtubule plus end before detaching\cite{yildiz2004kinesin}. One type of opto-cluster was formed by fusing an iLID domain and the associated micro-domain to the C-terminus of K401, forming K401-iLID and K401-micro~~\cite{Tyler2019}. When light activated, the homo-dimer K401-iLID binds to the homo-dimer K401-micro, creating a kinesin cluster that is able to generate interfilment microtubule sliding.  The second class of clusters, was formed using the 365 amino-acid fragment of the Drosophila kinesin-1 motor (K365)\cite{tayar2021active,chandrakar2022engineering}. K365 motors have a shorter neck linker domain and do not homo-dimerize. Additionally, K365 motors are monomeric and detach from a microtubule after each step~\cite{Berliner1995,HancockJOCB1998}. The iLID and micro domains were bound to the C-terminus of K365, forming K365-iLID and K365-micro, and under illumination they form a hetero-dimer. Notably, the opto-K365 motor clusters induce microtubule filament movement~\cite{LemmaL_arXiv_2022}. 

Opto-K401 and opto-K365 can power dynamics of active isotropic gels, which consist of an isotropic suspension of extensile microtubule bundles that generate turbulent flows ~\cite{LemmaL_arXiv_2022}. For opto-K401 the  maximum speed of the light-induced fluid flows decreased with time, starting at \SI{3}{\micro\meter\per\second} and slowing to \SI{1}{\micro\meter\per\second} over about $10$ hours. The fluid flow speed in the dark state (when the clusters were unbound) was low, but increased with time, ranging from \SI{5}{\nano\meter\per\second} to \SI{35}{\nano\meter\per\second} over $10$ hours. In contrast, for opto-K365 the maximum speed of the light-induced flows stayed constant with time, at  about \SI{1.2}{\micro\meter\per\second}. The speed was a function of light intensity, with a saturation speed at \SI{0.1}{\milli\watt\per\square\centi\meter}. The flow speed in the dark state was also nearly constant in time at a low speed of \SI{8}{\nano\meter\per\second}. The time it took for the speed to increase when illuminated and the time it took for the speed to decrease when illumination ceased were ~$10$ seconds and were independent of intensity.

\begin{figure}[h]
\centering
  \includegraphics[width=0.8\columnwidth]{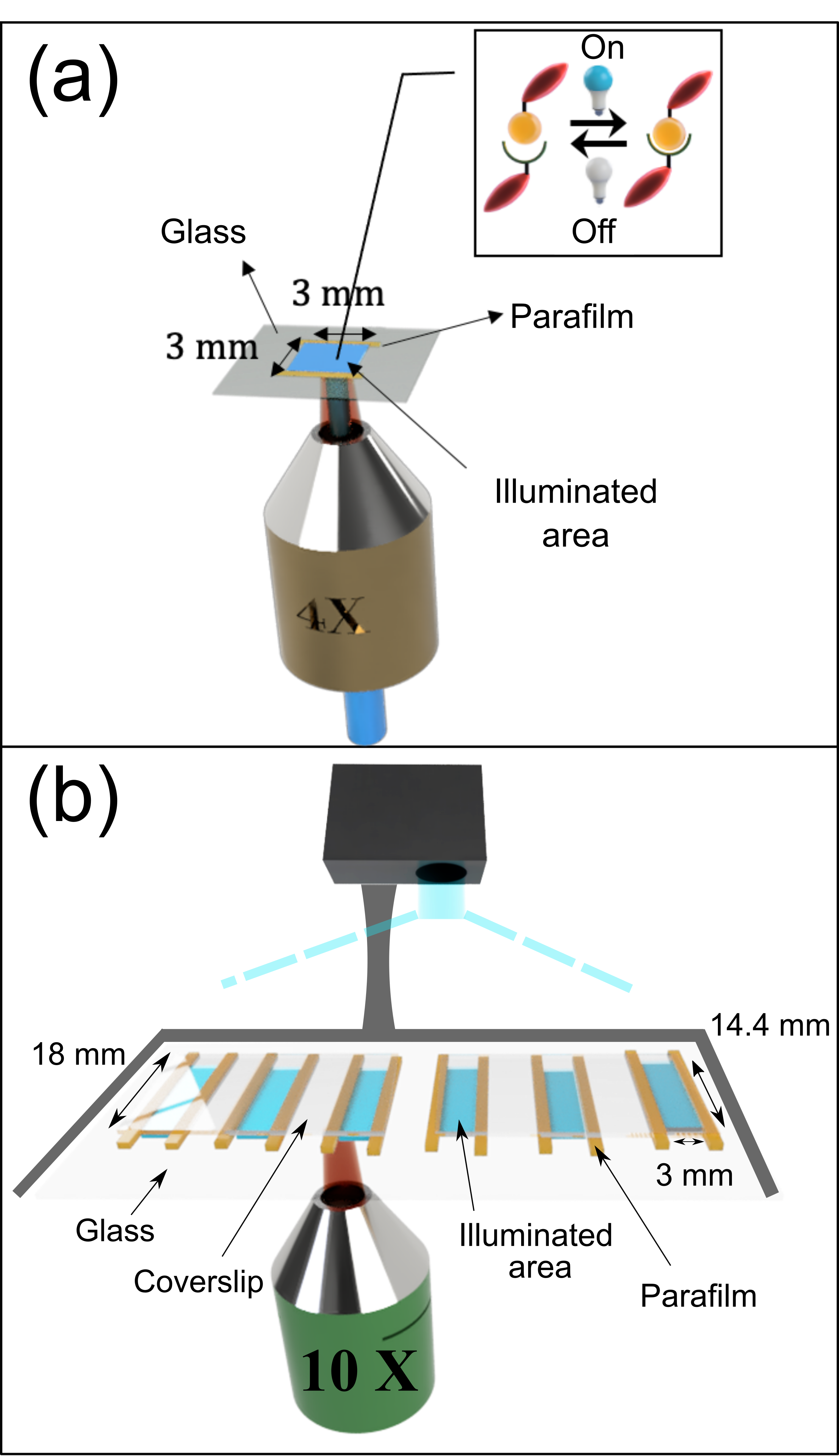}
  \caption{  Schematic of the experiment. (a) A single chamber of 3mm x 3mm was used for measurements of the flow speed of the active nematic. The same epi-illumination pathway  was used by the computer-controlled DLP projector to activate the nematic and for fluorescence imaging. (b) Up to six independent channels containing active nematics were employed to measure defect density. . The widths of each channel varied from 3mm to 6mm and the lengths were about 18 mm. The projector was mounted to the sample stage so the projected pattern (35 mm x 23 mm) translated with the sample as the sample was moved during imaging. }
  \label{fgr:SampleSchematic}
\end{figure}

\section{Materials and Methods}

\subsection{Kinesin light sensitive clusters}
We studied two designs of light-responsive kinesin pairs, opto-K401 and opto-K365, which were prepared as described previously in Lemma et. al.\cite{Tyler2019, LemmaL_arXiv_2022}. Briefly, plasmids for opto-K401 (K401-iLID, Addgene 122484, and K401-micro, Addgene 122485) and opto-K365 (K365-iLID and K365-micro) were transformed into BL21 cells. After culturing in 2XYTmedia, expression was induced at OD 0.6 with IPTG and grown for 16 hours at 18$^\circ$C. Cells were lysed with a tip sonicator and the lysate was clarified. Purification was performed using an AKTA Fast Protein Liquid Chromatography system (GE Healthcare) on a 1 mL Nickel column (HisTrap, GE Healthcare) and eluted in 500 mM imidazole, 50 mM sodium phosphate, 250 mM sodium chloride, 4 mM magnesium chloride, 0.05 mM ATP, 5mM mercaptoethanol, 5\% w/v glycerol, pH 7.2. The MBP tag on the K401-micro construct was digested with TEV protease and removed by passing the solution over a nickel column. The buffer was exchanged into 20 mM imidazole, 50 mM sodium phosphate, 250 mM NaCl, 4 mM \ce{MgCl2}, 0.05 mM adenosine triphosphate (ATP), 5mM beta-mercaptoethanol, 5\% w/v glycerol, pH 7.2 and then diluted to 50 percent v/v glycerol. The kinesin motors were aliquoted, flash-frozen in liquid nitrogen, and stored at -80$^\circ$C.

\subsection{Digital Light Projector (DLP)}
We utilize a Digital Light Processor (DLP) projector (EKB Technologies Ltd, EKB4500MKII P2) to project 460 nm wavelength light onto the sample. The specific DLP model is provided with a lens that casts a 20.4 mm by 12.7 mm image at a 25 mm working distance. By removing a stop in the telescopic lens and then shortening the working distance by a few millimeters, we generated a focused image of 35 mm x 23 mm. The DLP contains a Digital Micromirror Device (DMD) consisting of millions of mirrors that can be controlled independently to create a specific light pattern. Patterns are sent to the DLP through MATLAB. The light intensity is adjusted in two ways, by changing the current to the LED, or by varying the duty cycle frequency of the mirrors. The projector suffers from vignetting. To correct the vignetting, we used a bare CCD camera sensor placed in the sample plane to capture multiple images of the sample, which were then combined to form a composite image. To compensate for  vignetting, we used a lookup table to flatten the projected intensity. See Section 1, DLP calibration, and Supplementary Figures 1-4, in Supplementary Information for details.

\subsection{Sample cells, illumination and liquid crystal sample preparation}
Figure \ref{fgr:SampleSchematic} displays two different experimental geometries. The first geometry, Fig. \ref{fgr:SampleSchematic}(a), was primarily utilized to analyze the velocity field. In the case, the experimental chamber was fabricated by sandwiching two pieces of parafilm between a microscope slide and a coverslip. The flow cell had dimensions of \SI{3}{mm} length and \SI{3}{mm} width, with a height of approximately \SI{100}{um}. The bottom slide was treated with Aquapel, a commercially available hydrophobic product, while the coverslip was coated with a polyacrylamide brush to make it hydrophilic and non-adhesive.  The epi-illumination branch of a Nikon microscope was used for both the computer controlled DLP that provided the blue light source for opto-activation of the motors and the fluorescence illumination used for imaging. 

The second geometry, Fig. \ref{fgr:SampleSchematic}(b), was primarily used to analyze the defects in the nematic field. In this configuration, the glass treatment and chamber construction were similar but up to six independent channels containing active nematics were employed. These channels vary from \SI{3}{mm} to \SI{6}{mm} in width, with a length of \SI{18}{mm}. The computer-controlled projector provided spatio-temporal patterns of blue light used to activate the opto-kinesin clusters in the nematic. The DLP projector was mounted onto the sample stage and projected light from above onto the sample and into the objective located below the sample in a transmission illumination geometry. A separate optical arm in the reflection illumination geometry was used to illuminate and record the director using fluorescence microscopy. High resolution images of the entire sample were obtained by scanning the stage across the microscope objective. The projector was fixed to the sample stage so that as the stage moved, the pattern of projected light with respect to the sample did not change. More details are provided in the DLP calibration section of Supplementary Information.

To make the active mixture, we polymerized GMPCPP-stabilized microtubules labeled with AlexaFluor-647 dye\cite{tayar2022assembling,decamp}. These were combined with the opto-kinesin pairs (kinesin-iLID and kinesin-micro) at \SI{0.03}{\mg\per\ml} and reagents in concentrations described previously for active microtubule nematics: \SI{1.3}{\mg\per\ml} tubulin,  0.64\% [w/v] polyethylene glycol (\SI{35}{\kilo\dalton}), \SI{5.5}{\milli\molar} dithiothreitol, \SI{1.42}{\milli\molar} adenosine triphosphate, \SI{3.3}{\mg\per\ml} glucose, \SI{0.22}{\mg\per\ml} glucose  oxidase, \SI{0.039}{\mg\per\ml} catalase, \SI{3.28}{\milli\molar} magnesium chloride, \SI{26.7}{\milli\molar} phosphoenolpyruvate, and 2.8\% [v/v] pyruvate kinase/lactate dehydrogenase solution (Sigma P0294-5ML)\cite{tayar2022assembling}. 

For loading both types of experimental chambers, the channels were filled with a perfluorinated oil (HFE-7500) saturated with a fluoro-surfactant (RAN  Biotechnologies,  Beverly,  MA) at $5\%$  w/v. Then an active mixture was introduced with a pipette through one side of the cell while simultaneously removing the oil from the opposite side of the cell with a tissue. After filling, the cell was sealed with Norland 81 optical adhesive. This resulted in a flat, surfactant-covered oil-water interface.
The chamber was then exposed to \SI{20}{\milli\watt\per\square\centi\meter} (for the first geometry,  Figure \ref{fgr:SampleSchematic}a) or \SI{30}{\milli\watt\per\square\centi\meter} (for the second geometry, Figure \ref{fgr:SampleSchematic}b) of continuous blue light to create the nematic layer for 30 minutes. This causes the microtubules to bundle and sediment on the oil-water interface, forming the nematic.

\subsection{Data analysis}

Flow fields were obtained through particle image velocimetry (PIVLab MATLAB plugin)\cite{Pivlab,thielicke2014flapping}. The PIV analysis parameters were optimized for each data set. The window size was kept constant across data sets for each experiment. The average speed for each frame was calculated within the center of the chamber for even illumination. The spatial patterning experiment was employed to study how different blue light intensities affect the defect density in the active nematic. During the experiment, the defect distributions were identified using a home written MATLAB script\cite{MMN_Code} or by counting by hand.

\begin{figure*}
 \centering
 \includegraphics[width=14cm]{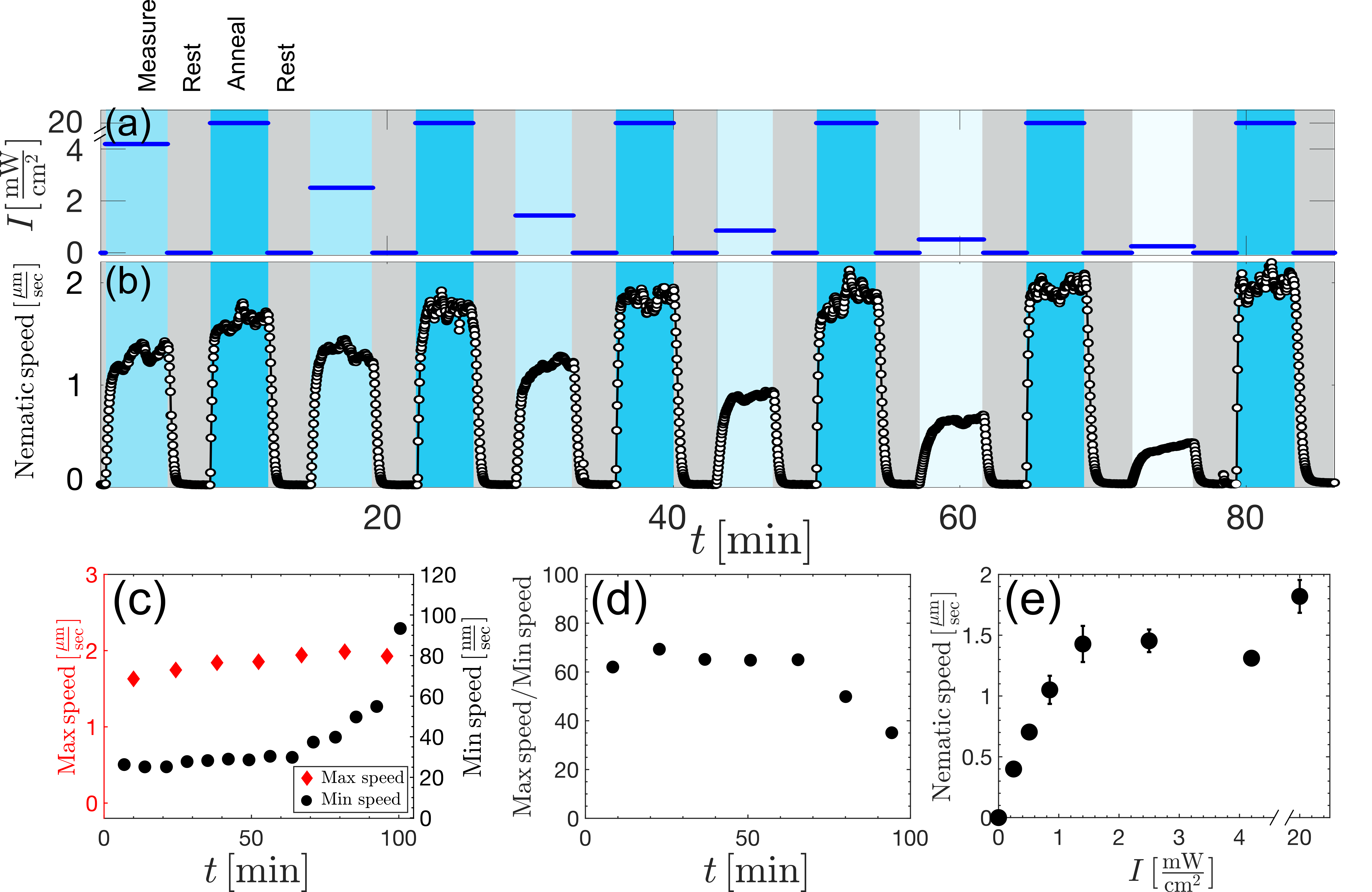}
  \caption{  Applied light determines active nematic flow via light-sensitive motor cluster crosslinking. (a) A DLP projector is used to apply spatially-uniform light onto an active nematic material composed of microtubules and light-sensitive motor clusters. The plot describes how the intensity of light (blue solid line), $I$, changes over time in accordance with the light-application protocol composed of 4 repeated steps: anneal at \SI{20}{\milli\watt\per\square\centi\meter} for 4 minutes, rest at \SI{0}{\milli\watt\per\square\centi\meter} for 3 minutes, measure at varied intensities for 4 minutes, and rest at \SI{0}{\milli\watt\per\square\centi\meter} for 3 minutes. These steps and the nematic response are shown in Supplementary Video 3. (b) The average nematic speed (white circles) is plotted against time for time-varying, spatially-uniform light intensities (described in (a)). When the applied light intensity is changed, the nematic responds, reaching steady-state speeds corresponding with applied light intensity. The average nematic speed was computed from the spatial average of the PIV flow field in the center of the sample cell shown in Fig. \ref{fgr:SampleSchematic}(a). (c) The maximum nematic speed (red diamonds, left y-axis) observed at saturating light intensity (\SI{20}{\milli\watt\per\square\centi\meter}) and the minimum speed (black circles, right y-axis) observed at \SI{0}{\milli\watt\per\square\centi\meter}  are plotted as a function of time. (d) The ratio of the maximum to minimum speeds as a function of time. (e) The steady-state speed of the active nematic as a function of applied light intensity.} 
  \label{fgr:NematicSpeedvsTimeProtocol}
\end{figure*}

\section{Experimental results}

\subsection{Demonstration of light-controlled 2D active nematic flow}
Kinesin motor clusters, formed through light-activated crosslinking, incorporated into active 2D MT nematics generating non-equilibrium dynamics. Homodimerized pairs of K401-iLID and K401-micro motors (opto-K401) were incorporated into depletion-induced 2D microtubule nematics. We observed the formation of a 2D layer of depletion-induced bundled microtubules undergoing active nematic average flow speeds of about \SI{3}{\micro\metre\per\second} when exposed to blue light (Supplemental Video 1). The nematic exhibited conventional topological defects of charge $+\frac{1}{2}$ and $-\frac{1}{2}$. These motile defects were continuously nucleating and annihilating. The defect-defect spacing was \SI{150}{\micro\meter}, similar to active nematics powered by conventional motor clusters. When the blue light was removed, the defect dynamics decreased but did not halt (Supplemental Video 2).

\subsection{Improving the range of light-activated flow in 2D active nematics through biochemical design}
Examining the dynamics of the active nematics powered by opto-K401, we measured flow speeds even in the dark state. Such ``dark speed'' strongly limited the range of light-activated flow. To quantify this behavior, we constructed a 2D active nematic from opto-K401 and illuminated the left half with sufficient intensity for the nematic to reach maximal speed while taking care to ensure the right half was not illuminated. Particle image velocimetry (PIV), revealed that ``dark speed'' was $\approx$\SI{500}{\nano\meter\per\second}. The ratio of speeds between the light and dark regions was $\approx$6 (Supplementary Figure 5). 

Experiments with optogenetic domains fused with the truncated kinesin-1 construct, K-365, whose shortened tail-domain eliminates endogenous kinesin aggregation, reduced dark flow speeds in 3D active fluids (Figure \ref{fgr:LemmaOptoKinesinSchematic})\cite{lemma2022spatiotemporal}. We examined the impact of these K365-iLID K365-micro motor pairs (opto-K365) within much denser 2D active nematic materials. Implementing opto-K365 significantly reduces the dark speeds and the ratio of the maximum and minimum speeds (Supplemental Video 3). We observed dark illumination speeds as low as \SI{25}{\nano\meter\per\second} and speed range ratios of about 65 (Fig. 3, (c) and (d)). This is an order-of-magnitude improvement with respect to the opto-K401 clusters. A shared challenge with both opto-clusters is their short lifetime. Opto-K401 active nematics under constant illumination of \SI{2}{\milli\watt\per\square\centi\meter}, which is the lowest intensity that produces the maximum speed, have a lifetime of ~1 hour. This lifetime is similar to the lifetime we observed with opto-K365 (about 1 hour). Typically, samples with conventionally-designed kinesin motor clusters can last 10-15 hours\cite{chandrakar2022engineering}.

\subsection{Characterizing light-responsive nematic flow}
Next, we examined the relation between applied light and the response of the 2D active nematic. This analysis showed the material responds consistently and reproducibly over a time of about one hour before aging effects deteriorate the ability of the applied light to control the material's active dynamics. 

\subsubsection{Light-responsive flow measurement protocol}
We measured the speed of the nematic as a function of light intensity. Although the protocol we designed does not eliminate hysteretic dynamics, it is intended to make the measurements as reproducible as possible by establishing a shared initial condition for all experiments. Before each measurement at a given intensity, we performed a two-step preparation protocol: ``anneal'' and ``rest.'' The anneal step consisted of illuminating the sample with a high, saturating intensity of \SI{20}{\milli\watt\per\square\centi\meter} for 4 minutes and the rest-step consisted of placing the system in the dark for 3 minutes. The rest step was long enough for the nematic flow to halt. The anneal step produced rapid motion of the nematic and rapid creation and annihilation of defects. This procedure ensured the initial condition of the nematic director field was statistically the same each time we measured the speed at the measured intensity, $I$ (Fig. \ref{fgr:NematicSpeedvsTimeProtocol}(a)). A data set acquired on a different day is presented in the Supplementary Figure. 6(a \& b). 

By following this light application protocol across a range of intensities, as demonstrated in Fig \ref{fgr:NematicSpeedvsTimeProtocol}(a), we measure how active nematic dynamics depends on the intensity of applied light (Fig. \ref{fgr:NematicSpeedvsTimeProtocol}(b) and Supplementary Video 3). By subsequently analyzing the PIV-derived flow fields, we extracted details about how light application affected the range of flow speeds, the sample aging, the appropriate range of activation light intensities, and the response of the flow speed to changes in the applied light. These parameters are critical for more advanced applications of light control in 2D active nematics. 

\subsubsection{Range of light-controlled flow speeds}
As our goal is to develop a 2D nematic whose flow can be prescribed by applied light intensity, a key metric is the contrast between the maximally- and minimally-flowing states. We quantify this through the ratio of the maximum to minimum speed of the nematic flow. The maximum speed has an average value of about \SI{2}{\micro\meter\per\second} and slightly increases over time (Fig. \ref{fgr:NematicSpeedvsTimeProtocol}c). The minimum speed is nearly constant at about \SI{25}{\nano\meter\per\second}  and then begins to increase after one hour. The ratio of the maximum to minimum speed is constant at about 65 for the first hour and then decreases until the system abruptly stops after about 2 hours (Fig. \ref{fgr:NematicSpeedvsTimeProtocol}d). The decrease in the ratio results from an increase in the minimum speed. An example taken on a different day is presented in Supplementary Figure. 6(c \& d). These measurements establish that \SI{460}{\nano\meter} illumination leads to a nearly linear rise in the steady-state flow speed which then saturates at more intense illumination (Figure \ref{fgr:NematicSpeedvsTimeProtocol}e). The saturation speed varied from day to day, ranging from \SIrange{1}{3}{\micro\meter\per\second}. To demonstrate reproducibility, an example taken on a different day is presented in the Supplementary Figure. 7. 

\subsubsection{Nematic lifetime and aging}
Biologically-derived materials used as model soft materials as they can suffer from aging effects such as oxidation and aggregation. Our results suggest aging only exhibits a strong effect in light-controlled microtubule 2D nematics after about 1 hour. Figure \ref{fgr:NematicSpeedvsTimeProtocol}(c) shows that the average dark flow speed remained constant for about 60 minutes before it began to increase from a minimum value of \SIrange{25}{100}{\nano\meter\per\second} at 100 minutes, at which time the sample ceased to move. Likewise, the maximum speed gradually increased in time, but only by a few percent. Thus, there is a duration of 60 minutes before the sample shows the effects of aging. As a reference timescale, the next sections characterize the time for the defect density and nematic speed to change in response to a change in light intensity.

\subsubsection{Characterizing transient flow response} 
\begin{figure}[h]
\centering
 \includegraphics[width=.75\columnwidth]{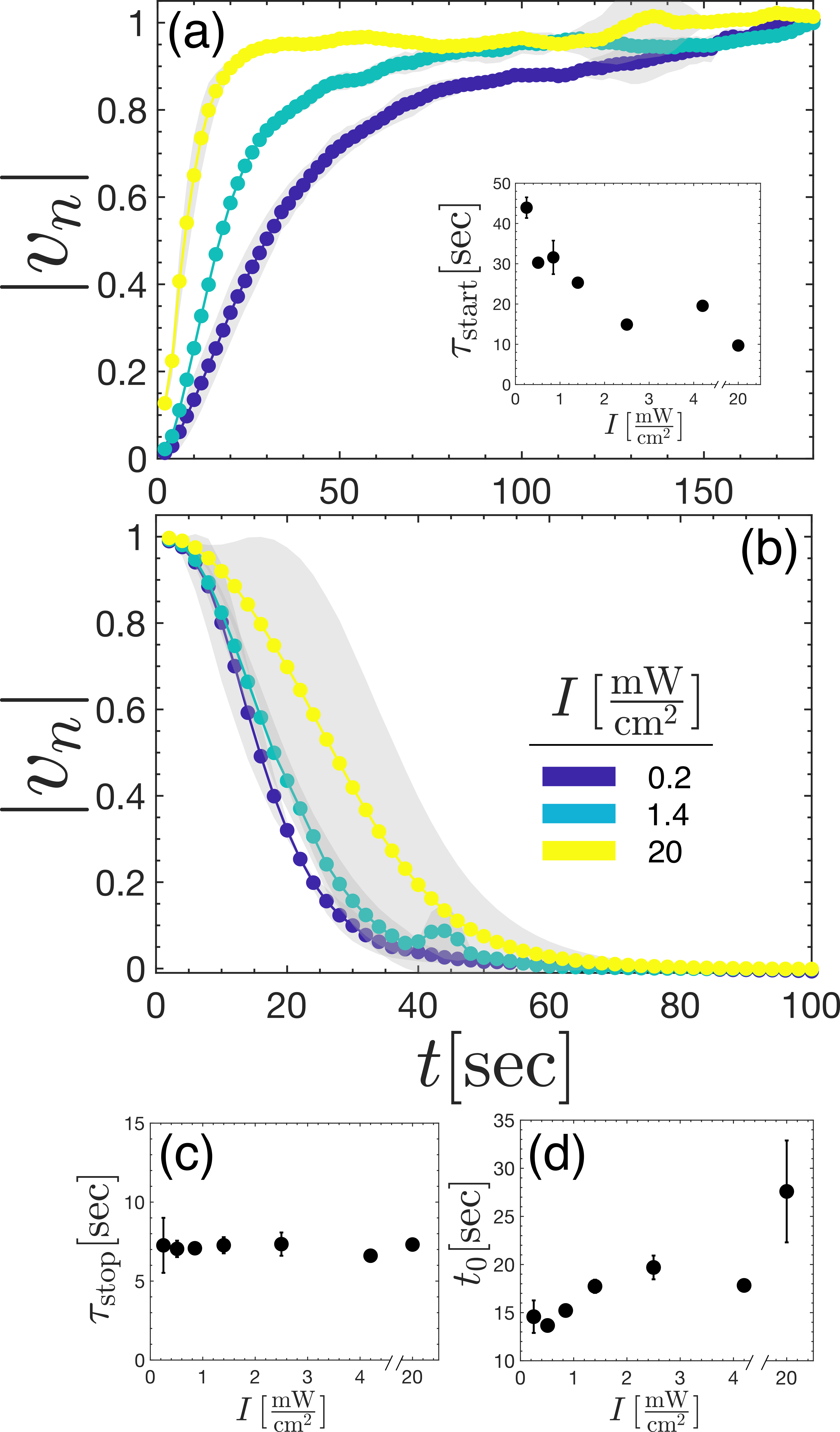}
  \caption{  Nematic flow exhibits transient response after changes in applied light intensity. (a) The applied light intensity is increased from \SI{0}{\milli\watt\per\square\centi\meter} to $I$. |$v_n(t)$| (the average speed instantaneous speed, |$v(t)$|, normalized by the steady-state speed, |$v_s(t)$|) increases as a function of elapsed time, $t$. The grey band represents the standard deviation. \textit{Inset}: The characteristic time constant, $\tau_{start}$, extracted from the function, $|v_n(t)|=1-\exp({-t/\tau_{\scriptsize \mbox{start}}})$, is plotted against light intensity, $I$. (b) The applied light intensity is decreased from $I$ to \SI{0}{\milli\watt\per\square\centi\meter}. The normalized speed, |$v_n$|, is plotted against elapsed time, $t$. (c) The data in (b) is fit to  $(1+\exp({(t-t_0)/\tau_{\scriptsize \mbox{stop}})})^{-1}$. Here, $\tau_{\scriptsize \mbox{stop}}$ is plotted against light intensity. (d) The time lag, $t_0$,  is plotted against the applied light intensity }
  \label{fgr:NematicStartStopRates}
\end{figure}

We established that the steady-state flow speed is determined by the magnitude of the applied activation light. As our goal is to control active nematics externally in real-time, it is necessary to characterize the transient response of the nematic flow to a light-induced change in the number of bound motor clusters. To determine the time constants which describe the transient flow response, we introduce the normalized speed, $|v_n(t)|=|v(t)|/|v_s(t)|$, where $|v(t)|$ is the instantaneous speed and $|v_s(t)|$ is the average steady-state speed. Fig. \ref{fgr:NematicStartStopRates}(a) plots the normalized speed against the time ($t$) following an increase in intensity from $0$ to $I$. After the applied light is increased, the resultant flow speed increases until saturation. The time it takes to saturate the flow speed decreases as the applied light intensity is increased. We fit this data to an exponential function, $|v_n(t)|^{{\scriptsize \mbox{start}}}=1-\exp({-t/\tau_{\scriptsize \mbox{start}}})$, and plot the time constant for the nematic to reach its steady-state speed after starting from rest, ($\tau_{\scriptsize \mbox{start}}$), as a function of the light intensity.
 
We plot the measured normalized speed against the time after the light is extinguished (Fig. \ref{fgr:NematicStartStopRates}(b)). Immediately after removing high-intensity illumination, the nematic speed remains constant (Fig. \ref{fgr:NematicStartStopRates}(b)). Only after a delay time, $t_0$, one observes a decrease in speed and at long times the sample attains its ``dark speed''. To quantify this process, the normalized speed is fitted to an exponential of the form, $|v_n(t)|^{{\scriptsize \mbox{stop}}}=(1+\exp({(t-t_0)/\tau_{\scriptsize \mbox{stop}})})^{-1}$. The function has two fitting parameters:  the time constant for the speed to decrease to zero, $\tau_{\scriptsize \mbox{stop}}$ (Fig. \ref{fgr:NematicStartStopRates}(c)),  and  the  time delay between extinguishing the illumination and when the speed begins to decrease, $t_0$ (Fig. \ref{fgr:NematicStartStopRates}(d)). Interestingly, $\tau_{\scriptsize \mbox{stop}}$ does not depend on light intensity. This is  expected as once the kinesin clusters unbind, the extensile forces will stop and we assume that the disassociation rate of a micro-iLID dimer unbinding into monomers is independent of the dimer concentration of dimers. To confirm the reproducibility, a data set acquired on a different day is presented in the Supplementary Figure. 8. 

  \label{fgr:StopRateSchematic}

\subsection{Patterning material structure with light}

\begin{figure}[h]
\centering
  \includegraphics[width=.8\columnwidth]{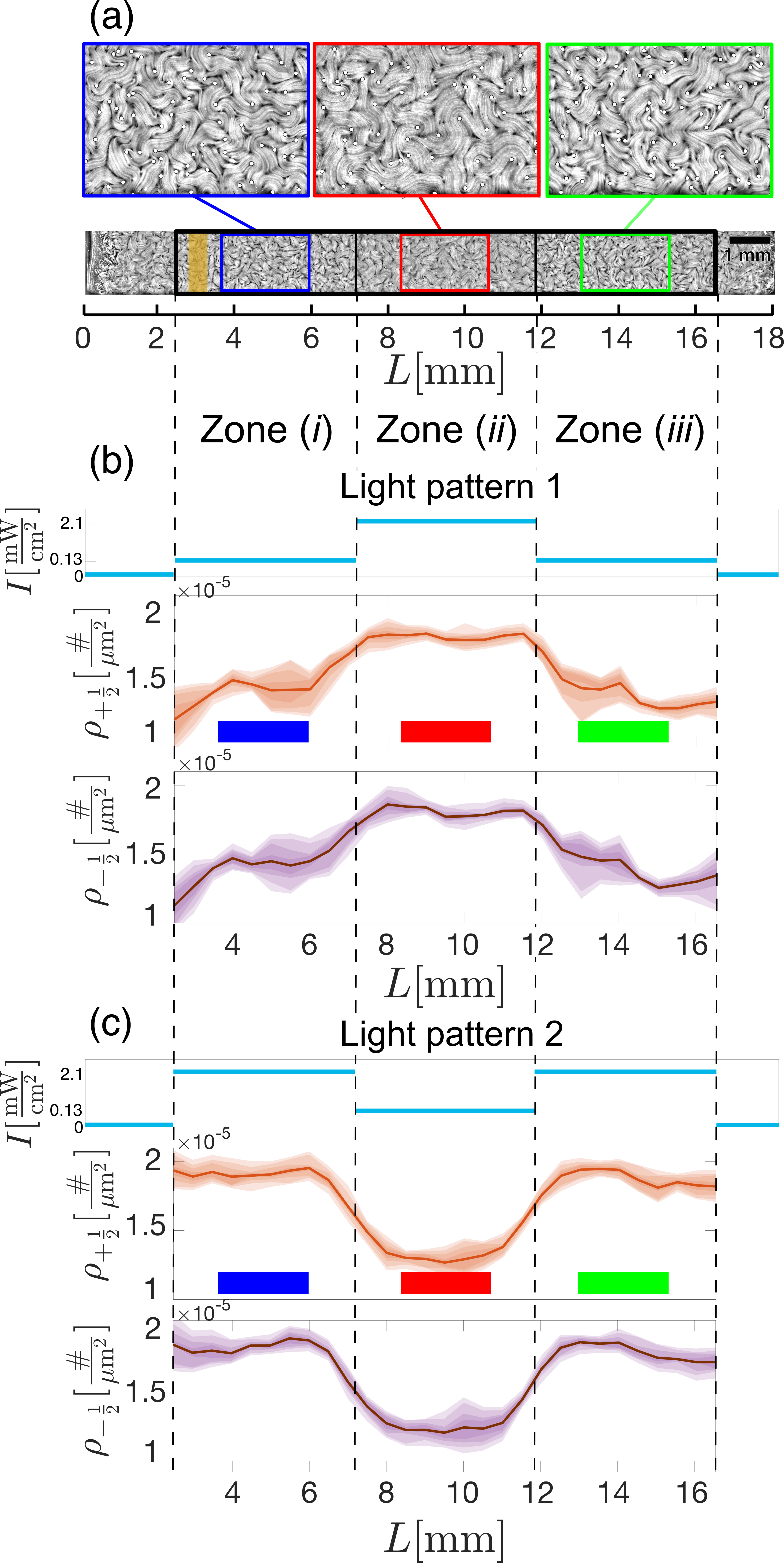}
 \caption{  Spatially-patterned light controls local defect density concentrations. \textbf{(a.)} A photograph montage of size 18 mm x 1.7 mm, constructed from 25 images of the nematic in from the center of a 18 x 5.5 mm$^2$ chamber (Fig. \ref{fgr:SampleSchematic}(b)). The three black rectangles indicated as zones (\emph{i}), (\emph{ii}) and (\emph{iii}) of area 4.8 mm x 1.7 mm are regions onto which blue light of uniform intensity is projected. The colored magnified regions of area 2.4 mm x 1.7 mm have the $+\frac{1}{2}$ defects indicated as white circles. \textbf{(b. \& c.)} The steady-state $\pm \frac{1}{2}$ defect densities ($\rho_{\pm \frac{1}{2}}$) are plotted against length ($L$) along the channel in panels (b) and (c).   Different light intensity ($I$) patterns were used in (b) and (c). \textbf{(d.)} The number of  $+\frac{1}{2}$ defects in each colored zone are plotted as defect density ($\rho_{+ \frac{1}{2}}$) vs. time (t).  The experiment began with the entire sample illuminated at \SI{30}{\milli\watt\per\square\centi\meter}. At t = 0, the intensity was changed to \SI{0.13}{\milli\watt\per\square\centi\meter} in the blue and green regions and \SI{2.1}{\milli\watt\per\square\centi\meter} in the red area.  Then, after 33 minutes (indicated by the right edge of the gray bar), the intensity was reversed, with \SI{2.1}{\milli\watt\per\square\centi\meter} in the blue and green regions and \SI{0.13}{\milli\watt\per\square\centi\meter} in the red region.}
  \label{fgr:DefectDensityIntensityTime}
\end{figure}

So far, we quantified how varied light intensities affect the dynamics of nematics driven by light-sensitive kinesin motor clusters. However, the structure of the active nematics also depends on the applied light. Next, we establish that spatially-heterogeneous light patterns induce spatially-patterned material structure in opto-K365 active nematics through the density of nematic defects. 

To measure the structural changes imposed by a spatially heterogeneous pattern of light, we constructed a large (\SI{18}{\milli\meter} by \SI{5.5}{\milli\meter}) experimental chamber (Fig. \ref{fgr:SampleSchematic}(b) and \ref{fgr:DefectDensityIntensityTime}(a)). This allows for the experimental chamber to be divided into three spatially-separated zones large enough to collect statistically significant measures of the defect densities while avoiding the effects of the chamber walls (Supplementary Figure 9 \& 10). Each zone is illuminated with a uniform light intensity, but the intensity within each zone differs. The different densities of defects between the spatially-heterogeneous light application zones show that light can be used to prescribe the structure of the material through the density of defects.

\subsubsection{Steady-state pattern of nematic defects}
Figure \ref{fgr:DefectDensityIntensityTime}(a) shows a composite image taken across the full experimental chamber while spatially-patterned light is directed onto the sample in three uniformly illuminated zones. The cropped images in Fig \ref{fgr:DefectDensityIntensityTime}(a) illustrate the changes in the defect structure. To quantify this structural change, we identified defects within the composite images and calculated the local defect density. The specific light pattern and the corresponding defect density are plotted in Fig. \ref{fgr:DefectDensityIntensityTime}(b) as a function of distance across the sample. The pattern of activation light was then changed. Fig. \ref{fgr:DefectDensityIntensityTime}(c) shows the new spatial and the resultant changes in the defect density across the material. Across all the measurements, the defect density increases with increasing intensity. Additionally, the spatially and temporally-separated zones with equal applied light intensity generated structures with equal defect densities within experimental error. Table \ref{tbl:SpeedDensity} shows the full defect densities we observed in steady-state as a function of applied light. The density increases by 40\% across the measured light intensities. We did another experiment and analyzed it to show the data are reproducible. The data are presented in Supplementary Information table S1.

\subsubsection{Scaling of defect density with nematic speed}

\begin{figure}[h]
\centering
  \includegraphics[width=0.8\columnwidth]{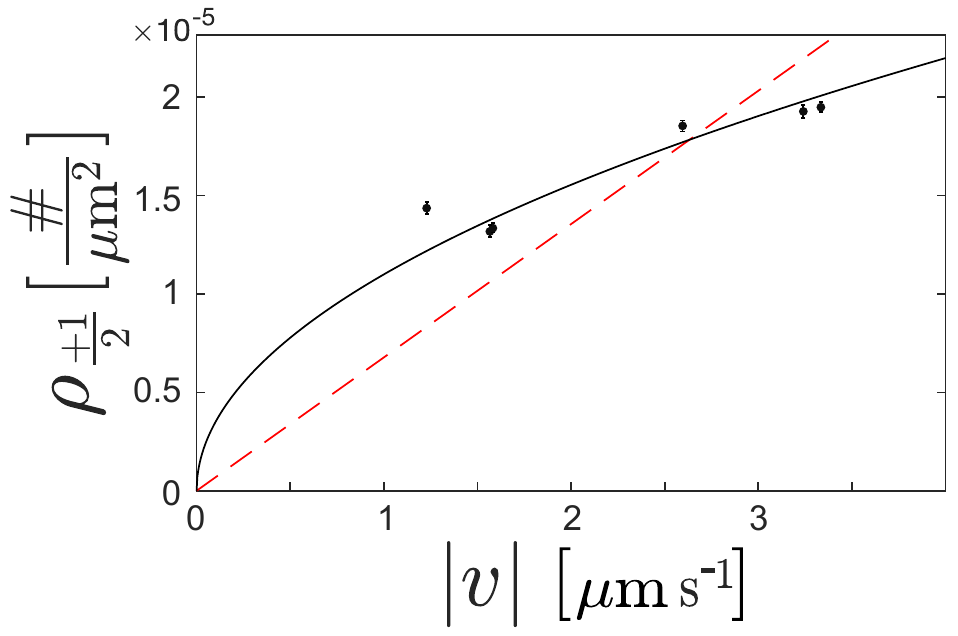}
  \caption{Defect density ($\rho_{+\frac{1}{2}}$) increases with the speed ($v$) of an active nematic (black circles). Fits are to $\rho \propto |v|^{1/2}$ (black solid line) and $\rho \propto |v|$ (red dashed line)}
  \label{fgr:DefectDensity_v_Speed}
\end{figure}

The structure and density of defects within active nematics have been extensively studied. Defect density is dependent on activity, $\alpha$, which is, in turn, proportional to the nematic speed: $|v| \propto \alpha$\cite{Thampi2014PhilTrans, Doostmohammadi2018}. Multiple theoretical models and experimental results find that the scaling behavior of defect density with activity $\alpha$ follows $\rho \propto \sqrt{\alpha}$ to $\rho \propto \alpha$\cite{Thampi2014PhilTrans, Doostmohammadi2018,hemingway2016correlation,VortexPaperLemma}. Figure \ref{fgr:DefectDensity_v_Speed} plots the steady-state defect density, $\rho(I)$, against the steady-state speed, $|v(I)|$. We have applied one-parameter fits which fix the value of $\rho$ to zero at $|v|=0$. The $\rho \propto \sqrt{\alpha}$ fit outperforms the linear fit, giving $R^2$ values of $0.86$ and $0.47$, respectively, in agreement with previous findings. Although this seems like strong confirmation, we note a paucity of data: the data for defect density varies only by a factor of 1.4 (Table \ref{tbl:SpeedDensity} and Supplemental Table 1) and we have fixed the zero-crossing in our fits. Furthermore, inside our sample chamber, the defect density from the cell boundary to the center varies by up to a factor of 4, as shown in Supplementary Figure 9 and Supplementary Figure 10. This gradient in defect density leads to a variation in defect density across the cell that is much greater than the change in the defect density induced by light. This large spatial gradient in defect density complicates experiments and, to minimize the impact of the gradient, all our measurements were taken in the center of the cell, where the gradient was at a minimum.

\subsubsection{Transient defect density}

\begin{figure}[h]
\centering
  \includegraphics[width=0.8\columnwidth]{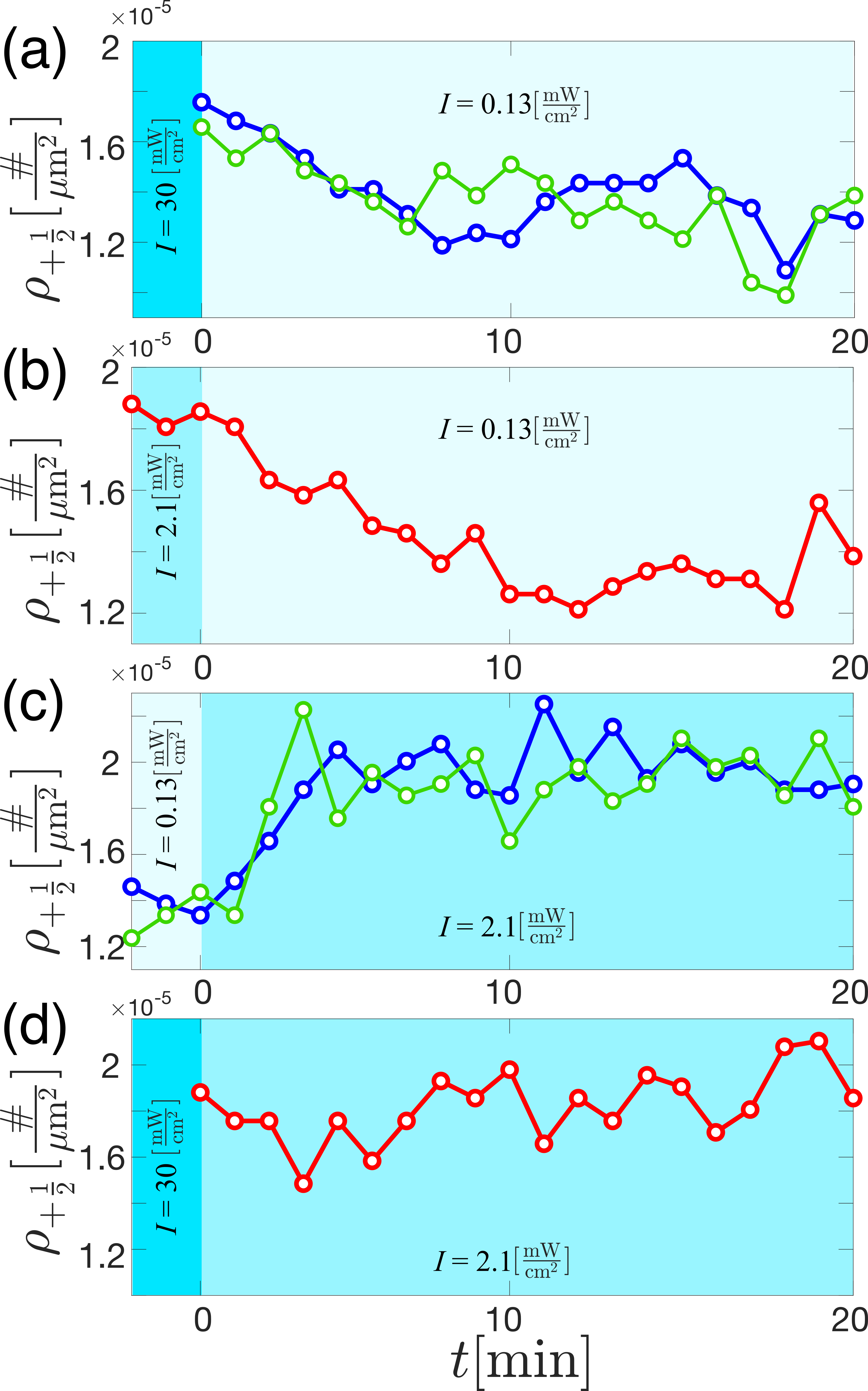}
  \caption{ Transitions to steady-state $+\frac{1}{2}$ defect density after changes in applied light intensity. (a) Defect density ($\rho_{+\frac{1}{2}}$) vs. time ($t$) as intensity is changed from \SIrange{30}{0.13}{\milli\watt\per\square\centi\meter} at $t$=0. (b) Defect density vs. time as intensity decreases from  \SIrange{2.1}{0.13}{\milli\watt\per\square\centi\meter}. (c) Defect density  vs. time when intensity is increases \SIrange{0.13}{2.1}{\milli\watt\per\square\centi\meter}. (d) Defect density  vs. time remains steady when intensity changes from \SIrange{30}{2.1}{\milli\watt\per\square\centi\meter}. }
  \label{fgr:DefectEvolutionOverTime}
\end{figure}

Next, we focus on how active nematics transition between two steady-state concentrations of defects. When the light intensity is increased, the defect density increases before reaching a steady-state value (Fig. \ref{fgr:DefectEvolutionOverTime}(a and c)). Analogously, when the light intensity is decreased (Fig. \ref{fgr:DefectEvolutionOverTime}(b)), the defect density decreases before reaching a steady state. Notably, the timescales for these transitions are from \SIrange{4}{10}{\min} while the steady-state flow speeds switch on times of less than \SI{50}{\second}. We also observe that, when the intensity is increased to a higher value, the transition time to the new steady-state defect configuration is faster. With increasing speed, the time for two defects to collide gets smaller and with increasing activity, the time to nucleate a defect decreases. This explains why the defect equilibration time with increasing light intensity is shorter than in the opposite case, as in Fig. \ref{fgr:DefectEvolutionOverTime}(b and c). Interestingly, Fig. \ref{fgr:DefectEvolutionOverTime}(d) shows no change in defect density over time. In this case, the intensity is changed from \SI{30}{\milli\watt\per\square\centi\meter} to \SI{2.1}{\milli\watt\per\square\centi\meter}. Both of these intensities are high enough that the flow speed is saturated. If the flow speed serves as a proxy for activity, then one expects the defect density to change only when the speed also changes. Indeed, this is the experimental observation.

\section{Discussion}
Light-sensitive motor clusters carry the promise of locally determining active stresses within active matter. In this work, we introduced experiments that show light-sensitive motor clusters create active 2D nematics whose defect structure and flow are determined by the intensity of light patterns through the activation of light-sensitive motor clusters. 

Our results also highlight the importance of microscopic biomolecular design in cytoskeletal materials. Similar to other recent work~\cite{LemmaL_arXiv_2022}, we found that non-processive opto-K365 reduced the dark flow of active microtubule materials. Interestingly, this improvement was more pronounced in 2D active nematics than for isotropic 3D active fluids. We don't have a full explanation of why this is the case. 

Conventionally-prepared active nematics experiments control the flow by changing the available concentration of the fuel substrate or density of permanently crosslinked motor proteins~\cite{Lemma2019}. Here, light-sensitive crosslinkers bind motor domains together upon illumination, which then create the microscopic stresses that lead to autonomous flow. As optogenetic domain binding is critical to the flow generation, it is not surprising that the flow response follows a behavior that is well-captured by chemical binding kinetics equations. As the blue light is turned on, the optogenetic crosslinkers will become activated with the magnitude of the light determining the fraction which can bind. The light intensity-dependent on-rate of the nematic flow suggests this is the underlying process controlling the onset of flow. Similarly, the independence of the applied intensity to the flow-stopping rate is suggestive that the halting of the nematic flow depends solely on the unbinding of crosslinkers. However, there remains a disconnect between the binding of motor clusters and how these microscopic constituents organize at the molecular level to produce the nematic flow which we observe and characterize. A better understanding of these mechanisms would help the rational design of improved experimental systems and broaden the range of phenomena we might observe.

The cessation of active nematic flow speed exhibited an unexpected lag time, $t_0$, before the speed starts to decrease (Fig. \ref{fgr:NematicStartStopRates}). We offer a hypothesis which relies on 5 assumptions (Supplementary Figure 11); (1) nematics moving with a steady-state speed contain kinesin motor clusters bridging anti-parallel aligned microtubule filaments that are exerting extensile force on the filaments, (2) it takes time for the iLID microdomains to fall apart after the removal of light, (3) the concentration of bound clusters increases with increasing intensity ($c_b(I)$), (4) above a critical concentration of bound domains the nematic speed remains constant ($c_v$), and (5) it is possible for $c_b(I)$ to be below or above $c_v$. These 5 assumptions explain that $t_0 = 0$ as long as the intensity is low enough so that the speed has not saturated, e.g. $c_b(I) < c_v$, because the speed reduces with each  cluster that breaks apart. However, when $c_b(I) > c_v$ the concentration of clusters must decrease to be less than $c_v$ in order for the speed to decrease and this takes time, so $t_0$ increases with intensity once the speed has reached its maximum.

A significant challenge remains in addressing the short lifetime of these experiments before the onset of aging effects (1 hour) with respect to the slow annealing time of defects (10 minutes), especially the annealing out of defects at slow speeds. Our experiments revealed that the active nematic aging was most apparent in the dark state speed, while the maximum speed under high-intensity illumination changed little over time. We expect this is the result of non-specific aggregation of motor proteins over time. However, the initial hour before  the onset of aging processes (such as increased dark flow speeds) is not accounted for by models in which the number of bound clusters is proportional to exposure and the speed increases with the number of bound clusters, suggesting the need for a more refined explanation. Diffusion chambers, which constantly restock substrate chemicals such as ATP, may help extend the sample lifetimes.

\begin{table}[h]
\small
  \caption{\ Speed and defect density as a function of intensity}
  \label{tbl:SpeedDensity}
  \begin{tabular*}{0.48\textwidth}{@{\extracolsep{\fill}}lll}
    \hline
    Intensity (\SI{}{\milli\watt\per\square\centi\meter}) & Speed (\SI{}{\micro\meter\per\second}) & Defect Density (\#/$\mu$m$^2$) \\
    \hline
    $2.1$ (High) & $3.05\pm0.34$ & $(1.91\pm0.1)\times10^{-05}$\\
    $0.13$ (Low) & $1.45\pm0.17$ & $(1.36\pm0.1)\times10^{-05}$ \\
    \hline
   & $\sqrt{\mathrm{ratio}}$ (High/Low)  $1.44$ & ratio (High/Low) $1.4$ \\
    \hline
    
  \end{tabular*}
\end{table}

\section{Conclusions}
Experiments with externally-defined activity give access to observations that force us to understand how self-organized steady-states emerge. Much of our understanding of 2D active nematics comes from experiments which probe steady-state self-organization. To directly observe the transient response of an active nematic switching between two steady-states, we introduced light-responsive motors which quickly change the activity of the material. These observations of non-steady-state structure and dynamics expose how active nematics move through configurational space from one dynamical state to another. Understanding how underlying physical mechanisms determine the pathways linking stable configurations is critical for introducing rationally-designed external control because it is precisely these non-steady-state self-organized states that we intend to engineer.

We assessed the feasibility of combining opto-K365 kinesin motors with extensile microtubule bundles to make light-controlled active nematics. We varied the speed of the nematic by a factor of 65, with a maximum speed  of \SI{2}{\micro\meter\per\second} and a minimum speed of \SI{25}{\nano\meter\per\second}. The time constant to increase the speed varied with intensity and was 44 seconds long at the lowest speed (intensity) and reduced to a few seconds at the highest speed (intensity). The time to decrease speed also depended on intensity, but was longest at 30 seconds for the highest intensity and reduced to 15 seconds at the lowest intensity. In summary, the nematic speed responded rapidly to large changes in intensity and we observed a large contrast between the minimum and maximum speeds.

The effect of light intensity on defect density ($\rho$) was smaller than the effect on speed ($v$) and we observed that $\rho \propto \sqrt{v}$. We measured the time constants for defect density to change upon increases and decreases in intensity and found that it takes about 10 minutes to decrease the defect density by 30\% and about 4 minutes to increase density by the same amount. It takes longer to decrease the defect density than to increase the defect density because when decreasing activity both the speed and the creation/annihilation rates of defects decreases. 

To achieve our goal of creating a light controllable active nematic, it is imperative to have samples that last much longer than the characteristic time to change defect density. As this time constant is of the range of 10 minutes, we estimate that samples with about a hundredfold longer lifetime are necessary to perform interesting control experiments. Consequently, creating nematics with substantially increased longevity is the most important experiment hurdle to creating fully controllable 2D active nematics. 

\section*{Author Contributions}
SF, JB and ZD conceptualized experiments. JB, LL and AH provided supervision. SF and JB wrote the paper. NS conducted initial experiments on K401. ZZ conducted the experiments on K365, data analysis and paper visualization. 


\section*{Conflicts of interest}
 There are no conflicts to declare.

\section*{Acknowledgements}
We acknowledge the Brandeis Bioinspired Soft Materials MRSEC-2011846





\bibliography{All_bib_Merged}
\bibliographystyle{rsc} 


\begin{titlepage}
    \centering
    {\scshape\Large Supplementary Information \par}
    \vspace{1cm}

\end{titlepage}

\onecolumn
\section*{Table of Contents}

\begin{enumerate}

\item DLP Calibration
\item Effect of the chamber walls on nematic speed and defect density
\item Transient nematic speed after a change in intensity 
\item Characterization of opto-K401
\item Gradients in defect density and nematic speed as a function of
distance from a boundary
\item Additional Supplementary Figures
\item Supplementary Videos
\end{enumerate}

\setcounter{section}{0}
\section{DLP Calibration} 

\subsection*{Flattening the light intensity}
In the light-activated active nematic experiment, the sample's illumination is achieved using a Digital Light Processor (DLP). The specific DLP model used is EKB4500MKII P2, produced by EKB Technologies Ltd which comes originally with a lens that gives a 20.4 by 12.7 image with a 25mm working distance. The size of the projected image can be increased by removing a limit screw from the zoom lens and decreasing the working distance by a few millemeters. This type of projector incorporates a Digital Micromirror Device (DMD) made up of numerous tiny mirrors that can be individually manipulated to generate a desired light pattern. The light pattern is transmitted to the DLP through MATLAB software and projected onto the sample.\\

However, after projecting a flat field onto the sample plane,  vignetting,   a reduction of the image's brightness  toward the periphery compared to the image center, occurs. In this experiment, it is essential to fix this problem because the motor protein is sensitive to light, and uneven illumination  leads to uneven activity.\\

To measure the extent of vignetting, the light intensity in the sample plane needs to be measured with a sensor first. The AVT Guppy F-146B camera was used for this experiment as a sensor. This camera is small, so it is easy to fit in the sample plane of the Nikon Eclipse microscope, easy to use, and easily modified to access the bare sensor. The bare sensor was then attached to a custom-made holder and mounted to the nosepiece on the Nikon microscope.(Supplementary Figure \ref{fig:SI_fig1}).\\

\begin{suppfigure}[htbp]
\centering
\includegraphics[width=10cm]{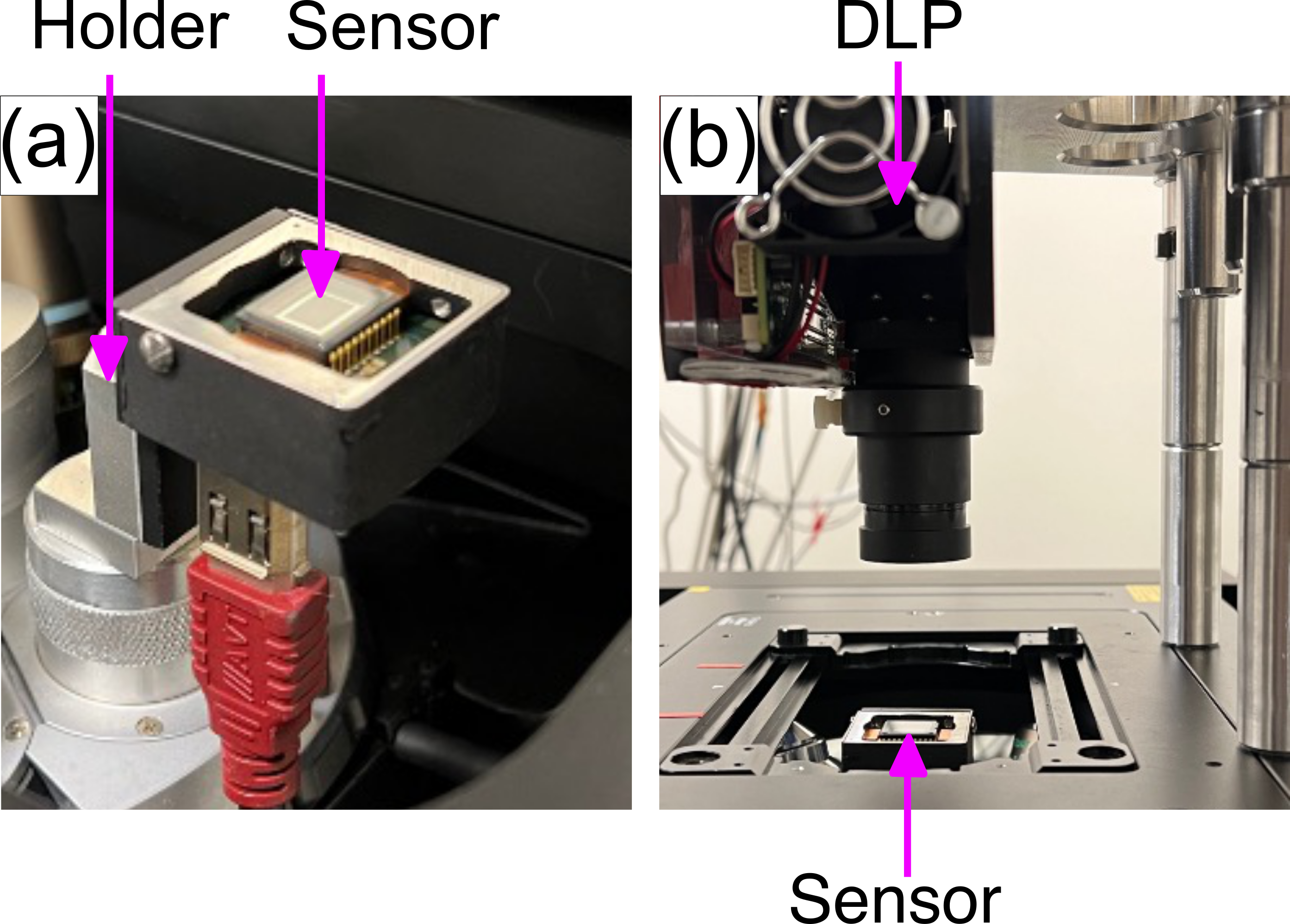}
\caption{ \textbf{: The sensor in the sample plane used to correct for vignetting in the DLP.} (a)The AVT Guppy F-146B camera is utilized for calibration. In order to access the sensor, the body of the camera is opened and as much of the housing is removed as possible. A custom machined holder is used to mount the sensor onto the Nikon nosepiece. The overall size of the sensor array is 4.65 mm by 4.65 mm.(b) The DLP is attached to the sample stage so that the projected light pattern moves with the stage. The sensor measures the light intensity projected by the DLP in the sample plane. The DLP used is EKB4500MKII P2 from EKB Technologies Ltd, and the wavelength is 460nm (center point).} 
 \label{fig:SI_fig1}
\end{suppfigure}

The camera sensor was too small to capture the entire projected field at once. To scan the entire projected field, the sensor was programmatically moved to different locations and multiple images were taken with 20 percent overlap and stitched together to create a composite image. This allowed for a detailed analysis of the light intensity across the sample and helped identify any areas of vignetting that needed to be corrected.
To improve accuracy, the 1 mm by 1 mm central part of the sensor is used.\\

The gaps between the mirrors  in the DMD  cause the measured intensity to fluctuate at a high spatial frequency. 
The DLP lens was defocused to smooth out the intensity fluctuations, but the resulting blurriness limited the sharpness of projected patterns (Supplementary Figure \ref{fig:SI_fig2}).
\begin{suppfigure}[htbp]
\centering
\includegraphics[width=8cm]{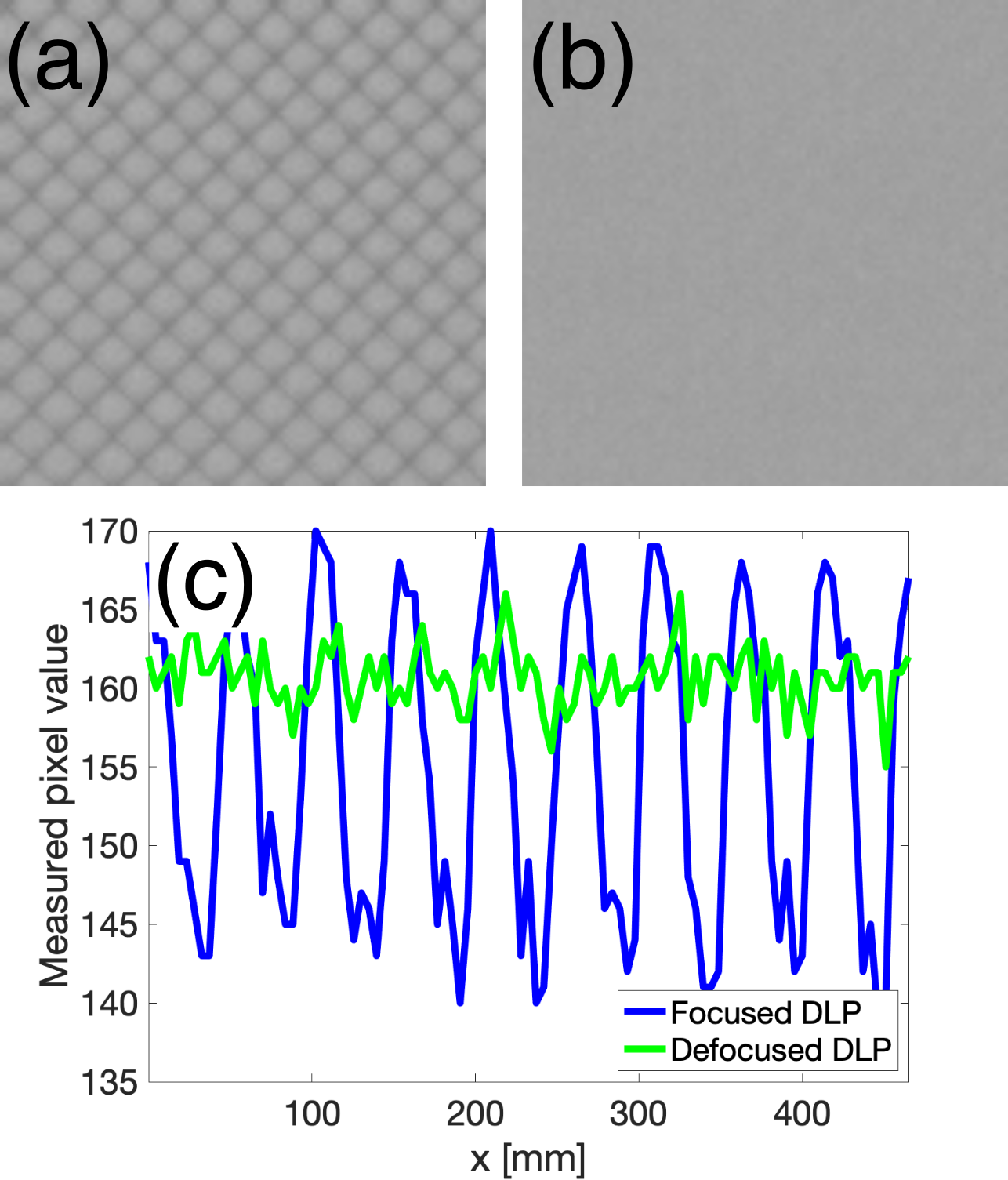}
\caption{\textbf{: The effect of defocusing the DLP lens.} (a) Focused lens, so each mirror is visible on the sample plane . (b) Defocused lens. (c) The intensity as a function of distance for the focused and defocused lens.} 
 \label{fig:SI_fig2}
\end{suppfigure}

We use a lookup table to correct for vignetting by creating a mapping between the projected and measured pixel values at each location as a function of the projected intensity. This mapping takes into account the nonlinear relationship between the projected and measured intensity values as a function of x-y position of the pixel on the DLP plane. Using this lookup table, the intensity values in the plane of the DLP can be selected to achieve the desired intensity for a given location in the sample plane (Supplementary Figure \ref{fig:SI_fig3}). This approach can be used to correct vignetting  (Supplementary Figure \ref{fig:SI_fig4}).

\begin{suppfigure}[htbp]
\centering
\includegraphics[width=10cm]{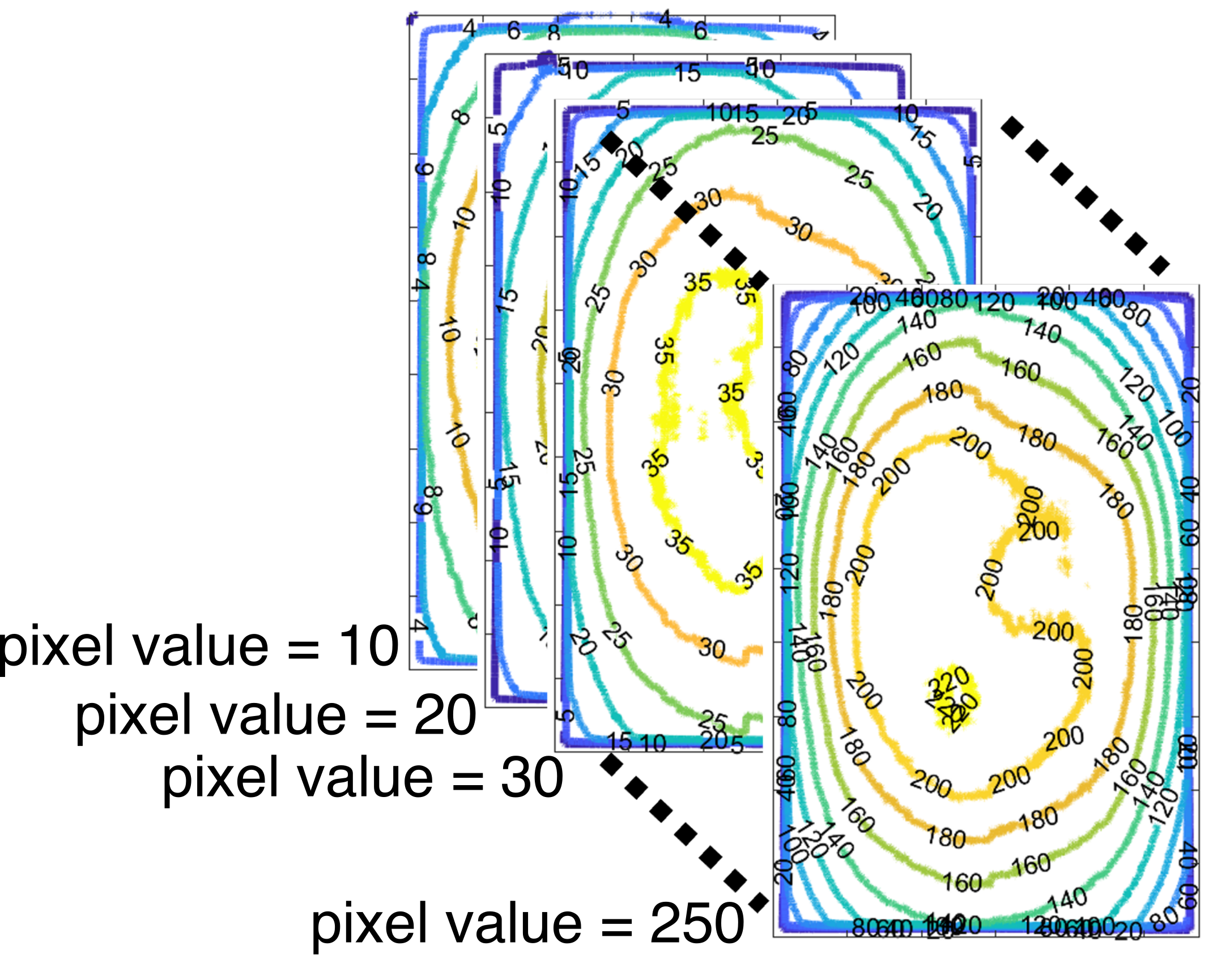}
\caption{\textbf{: Relation between projected light in the DLP plane and measured light in the sample plane.} A series of images in the sample plane were created by projected DLP light of constant intensity, varying in intensity ranging from 10 - 250 in steps of 10. Shown in the figure are the measured contour values of the projected intensity. Both projected and recorded intensities were 8-bit. The corresponding non-uniform sample-plane images were used to generate a lookup table relating projected pixel value to the measured intensity for each position in the sample plane.} 
 \label{fig:SI_fig3}
\end{suppfigure}

\begin{suppfigure}[htbp]
\centering
\includegraphics[width=11cm]{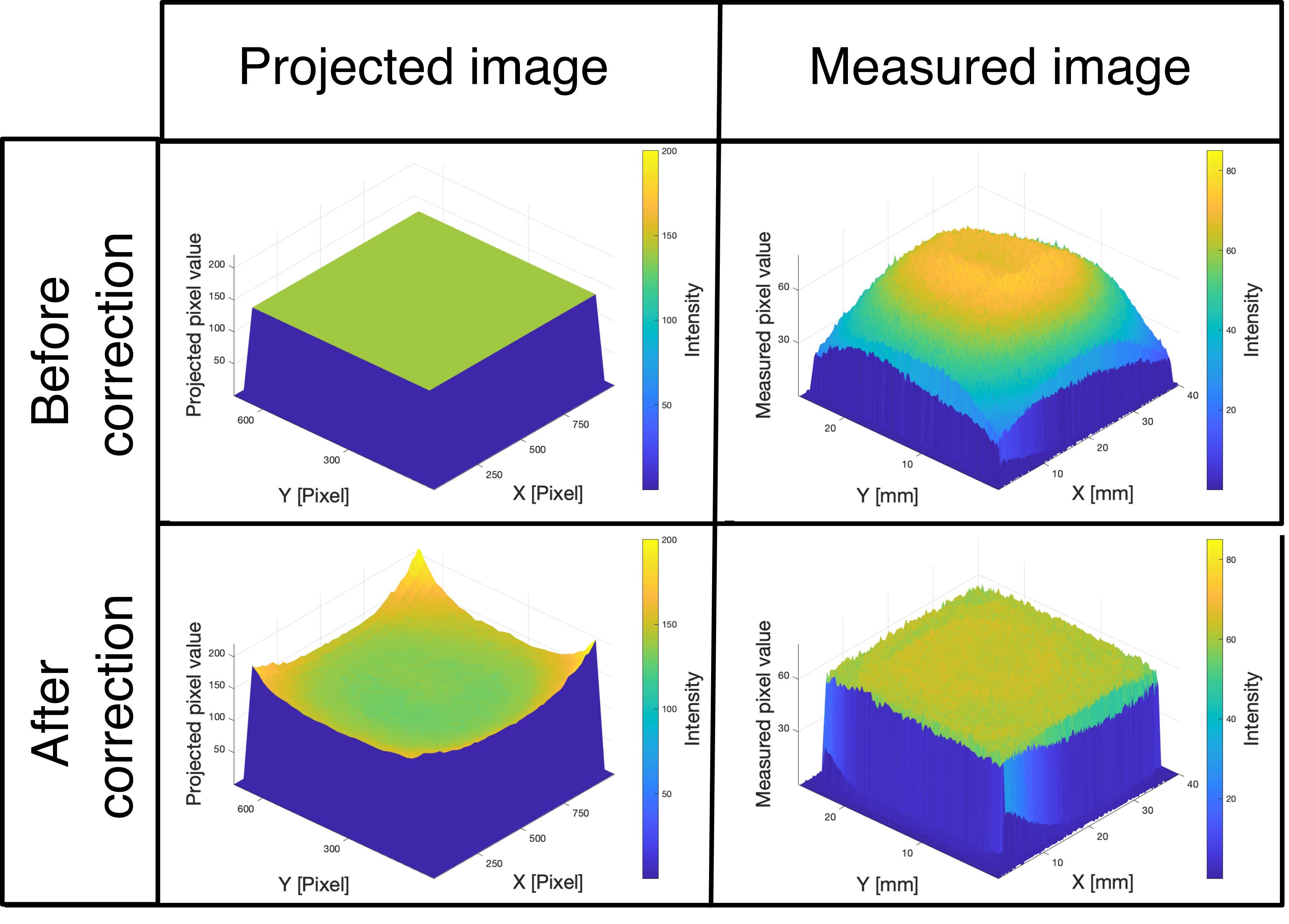}
\caption{\textbf{: Correcting for vignetting in the DLP.} The first row displays an image of uniform intensity, $I = 150$, in the plane of the DLP and the corresponding measured image recorded with a sensor in the sample plane. The area illuminated in the sample plane is 25mm x 40mm. Severe vignetting occurs. The second row shows an image of the intensity in the plane of the DLP after it has been processed to correct for vignetting, resulting in a more uniform image in the sample plane.} 
 \label{fig:SI_fig4}
\end{suppfigure}
\subsection*{Spatial Calibration}
Another critical aspect of using the DLP is the alignment of the field of view with the field of projection. The experimental setup consists of multiple channels that may require different intensities of light and so it is necessary to be able to relate the physical location of samples to the regions of the DLP that illuminate them. 

To ensure accurate alignment between the field of view and the field of projection, it is necessary to determine the transformation matrix that maps the coordinates of the projected image to the coordinates of the measured image. This can be done using a test pattern, which is first projected onto the sample plane and then measured with the microscope camera. The coordinates of the corners of the test pattern in the projected image and the corresponding coordinates in the measured image are used to determine the affine transformation matrix between the two coordinate systems.

Once the transformation matrix has been determined, it can be applied to the coordinates of each channel in the projected image to map them to the corresponding coordinates in the measured image. This allows for the precise borders of each channel to be defined, enabling the light to be accurately directed to the appropriate channels and ensuring that the correct intensities are applied to each channel.

It is important to carefully determine the transformation matrix to accurately align the field of view and projection. This may require some experimentation and fine-tuning to achieve the desired results. It is also essential to use a sufficiently complex test pattern that contains a sufficient number of corners or other distinctive features to determine the transformation matrix accurately.
The  transformation matrices that we employ are: 

\begin{equation}
\begin{bmatrix}
\cos\theta & -\sin\theta & 0\\
\sin\theta & \cos\theta & 0\\
0 & 0 & 1
\end{bmatrix}
\text{: Rotation}\quad 
\begin{bmatrix}
s & 0 & 0\\
0 & s & 0\\
0 & 0 & 1
\end{bmatrix}
\text{: Scaling}\quad 
\begin{bmatrix}
1 & 0 & t_x\\
0 & 1 & t_y\\
0 & 0 & 1
\end{bmatrix}
\text{: translation}\quad 
\end{equation}




\setcounter{section}{2}
\section{Effect of the chamber walls on nematic speed and defect density}
To study the effect of the chamber walls on nematic speed and defect density, we constructed a single large channel, 18 mm length and 5.3 mm width, using the experimental geometry shown Fig 2(b) in which the defect density and the nematic speed were measured across the channel. A zone of 12.8 mm in length by  5.3 mm in width was illuminated with blue light of 460 nm. This zone was divided into 3 contiguous areas of equal size, 5.3 mm in width and 4.3 mm in length, each with a different intensity, indicated in Supplementary Figure \ref{fig:SI_fig9} in cyan.


\setcounter{section}{3}
\section{Transient nematic speed after a change in intensity}
In Supplementary Figure \ref{fig:SI_fig6}, we measure the nematic speed as a function of intensity.
We followed the same two-step preparation strategy explained in the paper. Before measuring the speed at a given intensity, we performed a two-step preparation protocol: “anneal” and “rest.”  The anneal process involved lighting the sample at a high, sustaining intensity of 20  mW/cm$^2$ for 4 minutes, followed by a 3 minute rest period. This data shows the experiment's reproducibility shown in Fig 3 (a,b,c,d), and the corresponding nematic speed vs. intensity is shown in Supplementary Figure \ref{fig:SI_fig7} which confirms the reproducibility of Fig 3 (e).
 
Here, regions 1 and 3 were illuminated with 0.13 mW/cm$^{2}$, and region 2 was illuminated with 2.1 mW/cm$^{2}$. The defects vary in size near the walls, and the fluorescent image is low quality, so the MATLAB code \cite{MMN_Code} fails to find defects accurately. To solve this problem, the plotted defects are identified manually in supplementary Figure \ref{fig:SI_fig9} a,b,c panel (i, ii). The nematic speed in each region is calculated using PIV, and the average nematic speed is averaged 1.5 mm away from the walls, illustrated in Supplementary Figure \ref{fig:SI_fig9}d. To find the average defect density for each region, the defect density was identified using the MATLAB script\cite{MMN_Code} and averaged 1.5 mm away from the walls, shown in  Supplementary Figure \ref{fig:SI_fig9}e. As mentioned in the main manuscript, we changed the applied light intensity pattern after 33 minutes, therefore the 
regions 1 and 3 were illuminated with 2.1 mW/cm$^{2}$ and regions 2 was illuminated with 0.13 mW/cm$^{2}$. \\
 To determine the time constants which describe the transient flow response, we introduce the normalized speed $|v_n(t)|$ as the instantaneous speed $|v(t)|$ divided by the steady-state speed $|v_s(t)|$. Supplementary Figure. \ref{fig:SI_fig8}(a) plots the normalized speed ($|v(t)|/|v_s(t)|$) against the time ($t$) following an increase in  intensity from $0$ to $I$. This data shows that after the applied light is increased, the resultant flow speed increases until saturation. The time over which the sample takes to reach this saturated flow speed decreases as the applied light intensity is increased. We fit this data to an exponential $|v_n(t)|^{{\scriptsize \mbox{start}}}=1-\exp({-t/\tau_{\scriptsize \mbox{start}}})$ and plot the time constant for the nematic to reach its steady-state speed after starting from rest ($\tau_{\scriptsize \mbox{start}}$) as a function of the light intensity.

  In Supplementary Figure. \ref{fig:SI_fig8}(b), we plot the normalized speed against the time after the light is extinguished. To quantify this speed, the normalized speed is fit to an exponential of the form $|v_n(t)|^{{\scriptsize \mbox{stop}}}=(1+\exp({(t-t_0)/\tau_{\scriptsize \mbox{stop}})})^{-1}$, in which $\tau_{\scriptsize \mbox{stop}}$ is the time constant for the speed to decrease to zero and $t_{0}$ is the time delay between extinguishing the illumination and for the speed to begin to decrease.
  This data illustrates the reproducibility of the experiment shown in Fig 4.\\

\setcounter{section}{4}
\section{Characterization of opto-K401}
Supplementary Figure \ref{fig:SI_fig5} represents the active nematic powered by K401-iLID micro pairs in the dark and illuminated area. We constructed a 2D active nematic from K401-iLid micro pairs and illuminated half left, while half right was not illuminated.
To quantify the speed, we calculated the nematic speed in both regions using PIVlab \cite{Pivlab,thielicke2014flapping}, and we determined the nematic flow speed in the dark region was approximately 500 nm/sec. (Supplementary Video 1 for the illuminated and Supplementary Video 2 for the dark side) and the ratio of speeds between the light and dark regions was only a factor of six.

\setcounter{section}{5}
\section{Gradients in defect density and nematic speed as a function of distance from a boundary}
The averaged defect density and nematic speed at the center of the channel for the experiment explained in Supplementary Figure \ref{fig:SI_fig9} over both light patterns are presented in Table \ref{tbl:SpeedDensity}.

\begin{table}[htbp]
 \renewcommand\thetable{S 1}
\small
  \caption{\ Speed and defect density as a function of intensity}
  \label{tbl:SpeedDensity}
  \begin{tabular*}{1\textwidth}{@{\extracolsep{\fill}}lll}
    \hline
    Intensity (mW/cm$^2$) & Speed ($\mu$m/sec) & Defect Density (\#/$\mu$m$^2$) \\
    \hline
    $2.1$ (High)& $2.76\pm 0.38$ & $(2.67\pm 0.1)\times10^{-05}$\\
    $0.13$ (Low) & $1.19\pm 0.36$ & $(1.72\pm 0.18)\times10^{-05}$  \\
    \hline
   \hline
   & $\sqrt{\mathrm{ratio}}$ (High/Low)  1.51 & ratio (High/Low) 1.54 \\
    \hline
    
  \end{tabular*}
\end{table}

The reason for the spatial gradient in the defect density in Supplementary Figure \ref{fig:SI_fig9} panel (i)  is unclear. One possibility is that the oil gradient near the boundaries may cause the gradient. We made a 3 mm width and 3 mm length chamber using experimental geometry Fig 2(a). We captured multiple images using Confocal microscopy at different depths in the sample close to the wall to quantify the oil gradient. The nematic layer showed the location of the oil surface, and by analyzing the series of images, we could identify where the oil layer was positioned. To account for the oil's index of refraction, which is 1.29, the measured height was divided by this value. The point where the nematic layer is uniform was used as the reference point for the oil height axis. The result is illustrated in Supplementary Figure \ref{fig:SI_fig10}. There is a separation in length scales between the change in oil height from the boundary and the change in defect density. The oil height reaches a plateau value at 50 $\mu$m from the wall while the defect density plateaus at 1000 $\mu$m from the wall. Because of this separation in length scales we do not believe that the oil height causes the gradient in the defect density. 
\newpage

\setcounter{section}{6}
\section{Additional Supplementary Figures} 

\begin{suppfigure}[htbp]
\centering
\includegraphics[width=13cm]{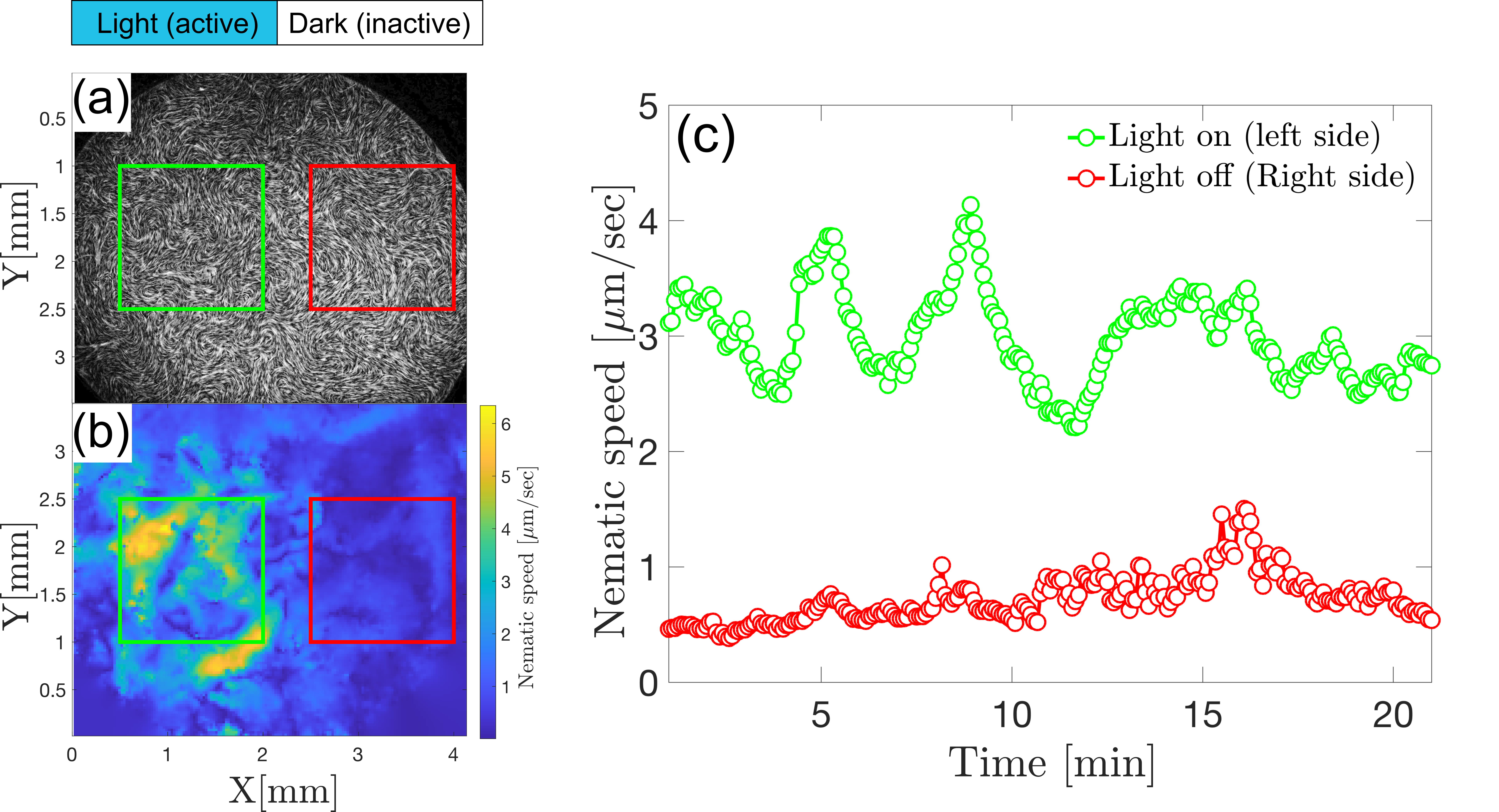}
\caption{\textbf{: Speed of K401 active nematic.} (a) Fluorescence micrograph of an active nematic of opto-K401. The exposure time was 100 msec. The left half was
illuminated with sufficient intensity to saturate the velocity and the right
half was dark. (b) Speed of the nematic as a function of position at an instant of time as determined by
PIV analysis. The exposure time was 100 msec. The speed is spatially
inhomogeneous. (c) Speed averaged over the regions within the color-coded boxes as a function of time. The speed fluctuates in time. The dark state of opto-K401 had a non-zero velocity and the average ratio of the velocity in the light (active) to dark (inactive) state was about 6.} 
 \label{fig:SI_fig5}
\end{suppfigure}

\begin{suppfigure}[htbp]
\centering
\includegraphics[width=10cm]{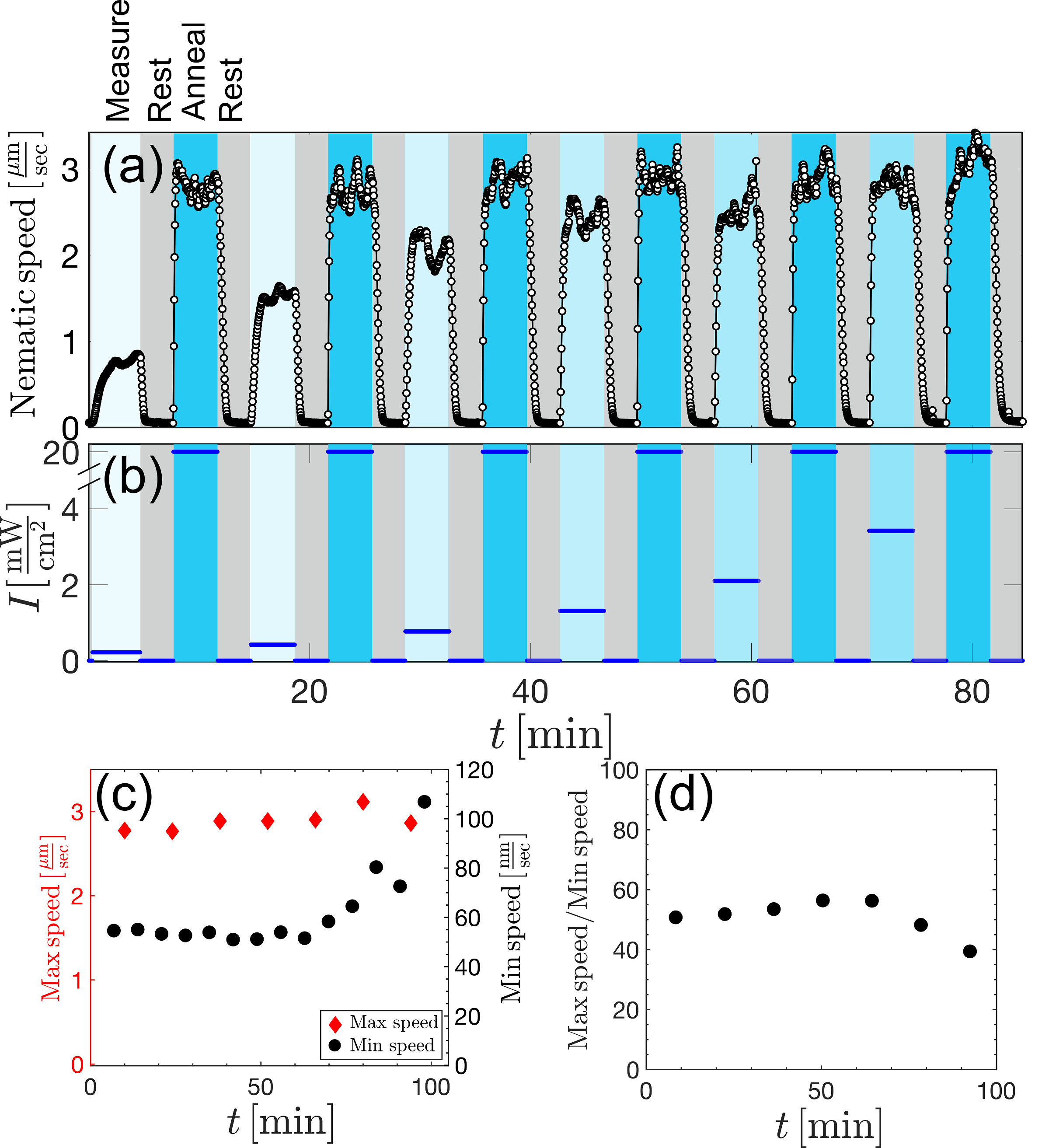}
   \caption{: Applied light determines active nematic flow via light-sensitive motor cluster crosslinking. (a) The average nematic speed (white circles) is plotted against time for time-varying, spatially-uniform light intensities. When the applied light intensity is changed, the nematic responds, reaching steady-state speeds corresponding with the applied light intensity. The average nematic speed was computed from the spatial average of the PIV flow field in the center of the sample cell shown in Fig. 2(a)(b) A DLP projector is used to apply spatially-uniform light onto an active nematic material composed of microtubules and light-sensitive motor clusters. The plot describes how the intensity of light (blue solid line), $I$, changes over time in accordance with the light-application protocol composed of 4 repeated steps: anneal at 20 mW/cm$^{2}$ for 4 minutes, rest at 0 mW/cm$^{2}$ for 3 minutes, measure at varied intensities for 4 minutes, and rest at 0 mW/cm$^{2}$ for 3 minutes. (c) The maximum nematic speed (red diamonds, left y-axis) observed at saturating light intensity (20 mW/cm$^{2}$) and the minimum speed (black circles, right y-axis) observed at 0 mW/cm$^{2}$ are plotted as a function of time. (d) The ratio of the maximum to minimum speeds as a function of time.} 
 \label{fig:SI_fig6}
\end{suppfigure}

\begin{suppfigure}[htbp]
\centering
\includegraphics[width=8cm]{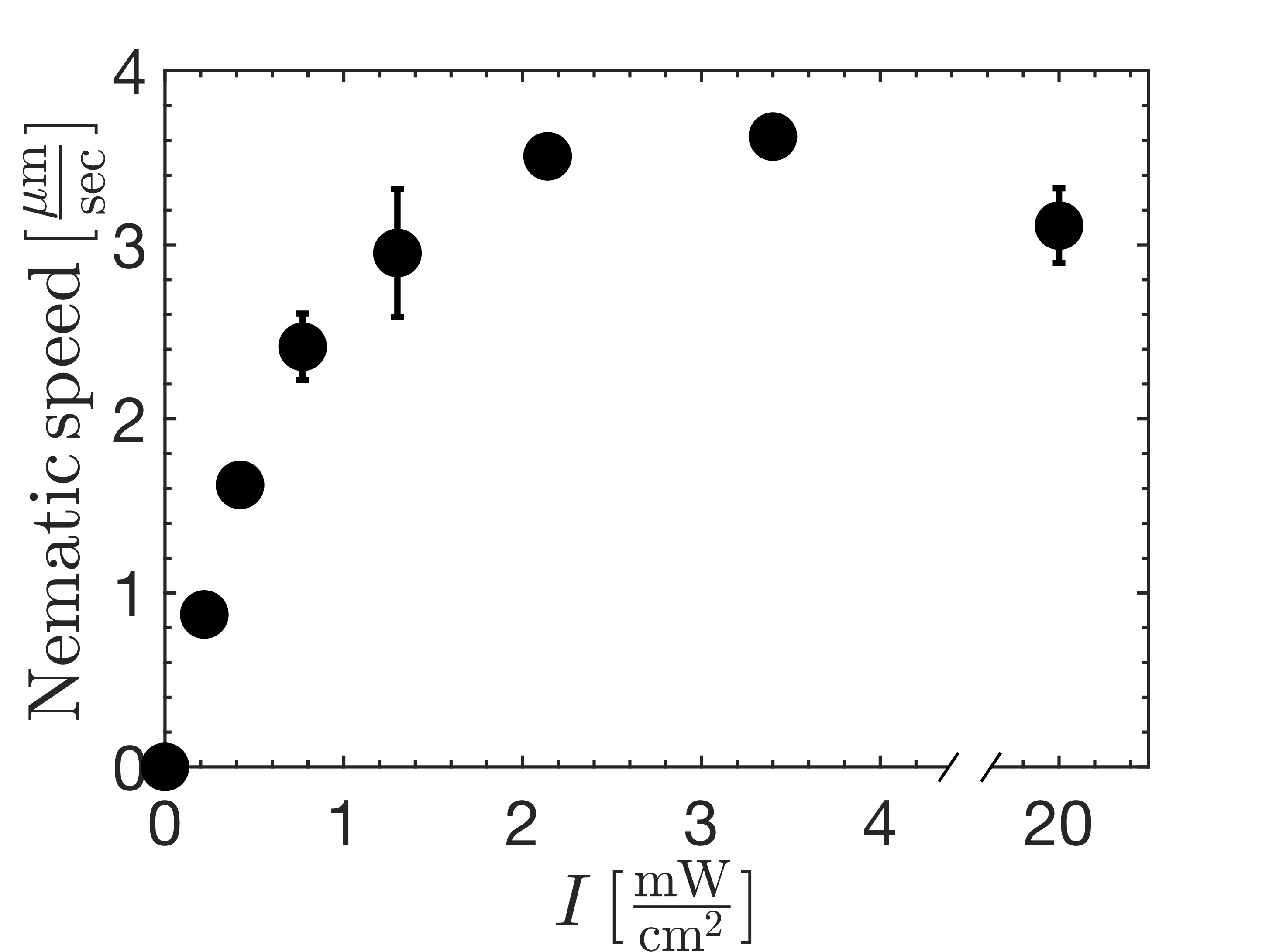}
  \caption{: Speed of opto-K365 active nematic vs. light intensity. Continuous illumination was employed and the speed at steady state was measured using PIV. The data was obtained in the center of the sample cell shown in Fig. 2(a).}
 \label{fig:SI_fig7}
\end{suppfigure}

\begin{suppfigure}[htbp]
\centering
\includegraphics[width=7cm]{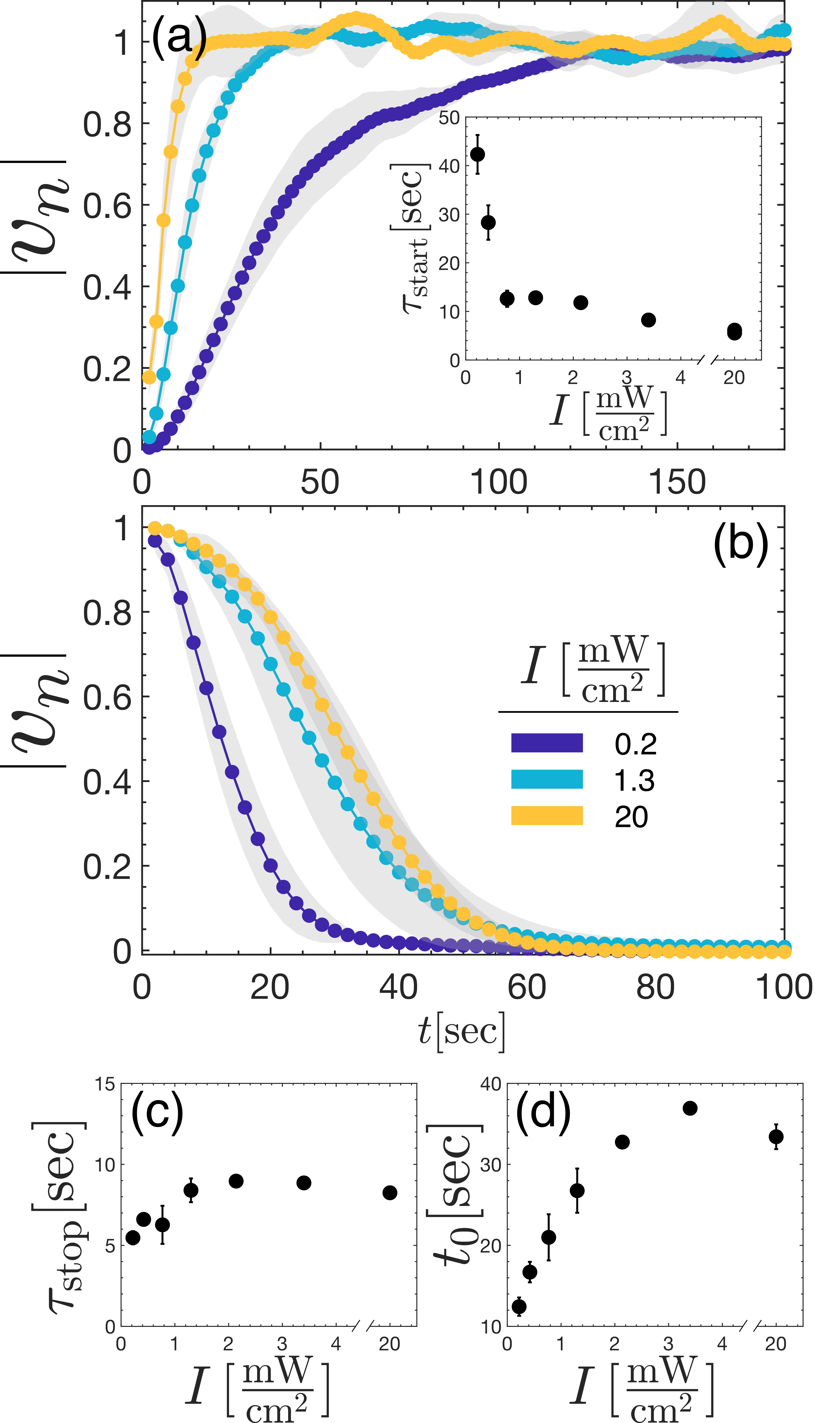}
  
    \caption{: Nematic flow exhibits transient flow after applied light intensity is changed. (a) The applied light intensity is increased from $0$ mW/cm$^{2}$ to $I$. |$v_n(t)$| (the average speed instantaneous speed, |$v(t)$|, normalized by the steady-state speed, |$v_s(t)$|) increases as a function of elapsed time, $t$. The grey band represents the standard deviation. \textit{Inset}: The characteristic time constant, $\tau_{\scriptsize \mbox{start}}$, extracted from the function, $|v_n(t)|=1-\exp({-t/\tau_{\scriptsize \mbox{start}}})$, is plotted against light intensity, $I$. (b) The applied light intensity is decreased from $I$ to $0$ mW/cm$^{2}$. The normalized speed, |$v_n$|, is plotted against elapsed time, $t$. (c) The data in (b) is fit to  $(1+\exp({(t-t_0)/\tau_{\scriptsize \mbox{stop}})})^{-1}$. Here, $\tau_{\scriptsize \mbox{stop}}$ is plotted against light intensity. (d) The time lag, $t_0$,  is plotted against the applied light intensity }
 \label{fig:SI_fig8}
\end{suppfigure}
\clearpage
\newpage

\begin{suppfigure}[htbp]
\centering
\includegraphics[width=14cm]{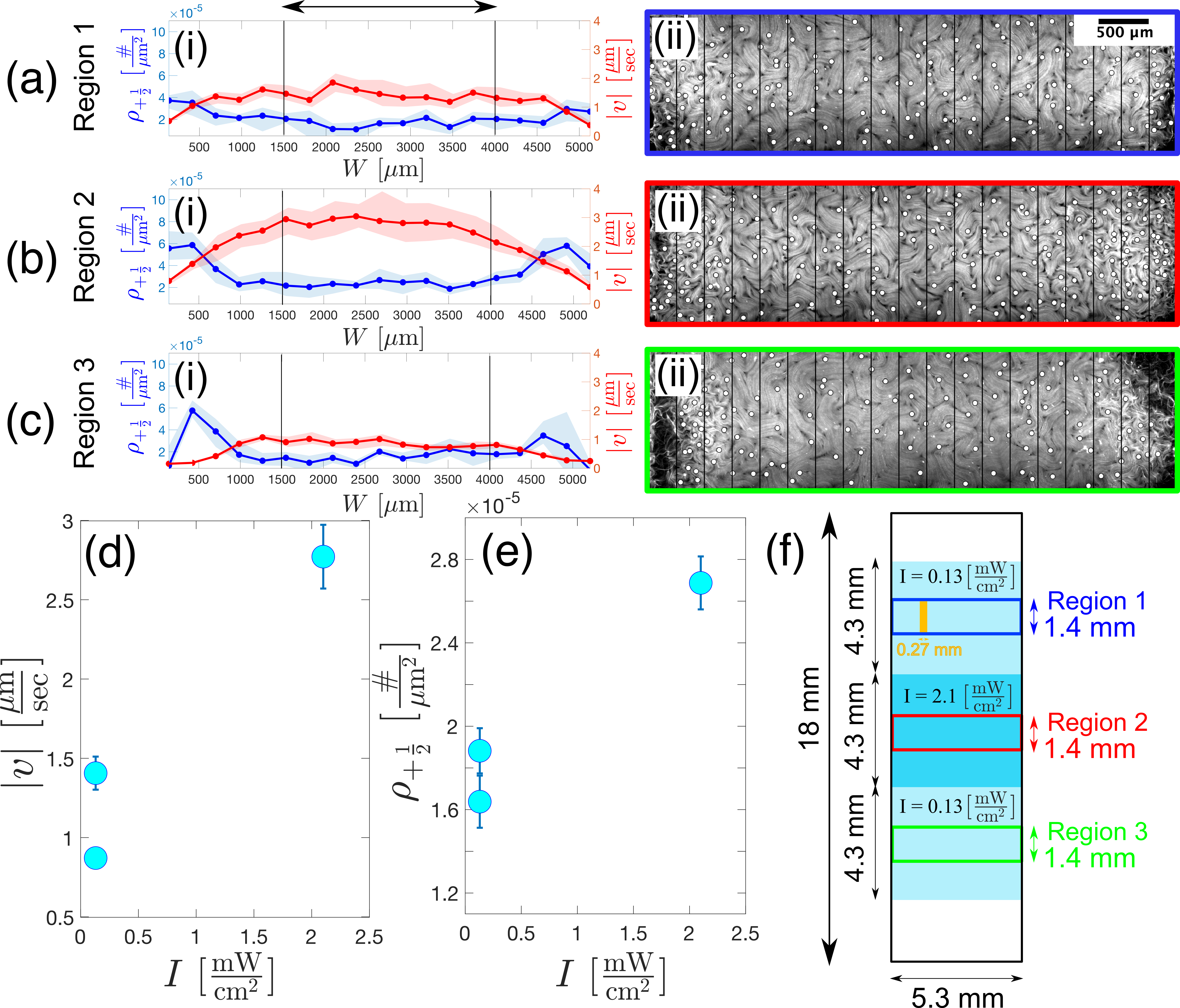}
\caption{ \textbf{Nematic speed ($v$) and defect density ($\rho_{+\frac{1}{2}}$) across the channel width ($W$).} \textbf{(a-c) panel (i)} The $+\frac{1}{2}$ defect density in blue ($\rho_{+\frac{1}{2}}$) and nematic speed ($v$) in red across the channel width ($W$) exposed to blue light in three different regions. The intensity ($I$) was $I = 0.13$ mW/cm$^{2}$ in Region 1 and Region 3, and  $I = 2.1$ mW/cm$^{2}$ in Region 2. The defect density is high on the channel edge and low at the channel center, while the nematic speed is low near the walls at the channel edge and high away from the walls at the channel center. Each data point in the plots in panel (i) corresponds to averages of $\rho_{+\frac{1}{2}}$ or $v$ in one of the black rectangles (1.4 mm length by 0.27 mm width) in panel (ii). \textbf{(a-c) panel (ii)} The corresponding experimental snapshots of the light-activated active nematic in 3 regions, each photograph is 5.3 mm width and 1.4 mm length, a montage of 6 images,  and each small black rectangle is 0.27 mm in width and 1.4 mm in length.  The white circles show the $+\frac{1}{2}$ defects. Scale bar, 500 $\mu$m.  The measurements are taken every 4 seconds, and the defects are identified manually. (d) The average nematic speed is calculated using PIV beginning at a distance 1.5 mm from the walls to avoid wall effects. The data included in the average speed is indicated in (a-c) panel (i) as those points between the two vertical black lines in the center of the channel and is from an area of 2.3 mm width by 1.4 length.   (e) The average $+\frac{1}{2}$ defect density for each intensity is measured 1.5 mm from the walls. The defects are identified using the MATLAB script mentioned previously. The data included in the defect density is indicated in (a-c) panel (i) as those points between the two vertical black lines in the center of the channel and is from an area of 2.3 mm width by 1.4 length.  (f) The entire chamber is  5.3 mm in width by 18 mm in length. The full illuminated area is 5.3 mm in width and 12.8 mm in length, which is divided into 3 areas of different intensities indicated in shades of cyan. Each illuminated area represented in a single cyan rectangle is 5.3 mm in width and 4.3 mm in length. The colored regions indicated as regions 1, 2 and 3 are 5.3 mm in width and 1.4 mm in length. The small orange rectangle of 0.27 mm in width and 1.4 mm in length represents the area over which the data in panel (i) is plotted.} 
 \label{fig:SI_fig9}
\end{suppfigure}

\clearpage

\begin{suppfigure}[htbp]
\centering
\includegraphics[width=7cm]{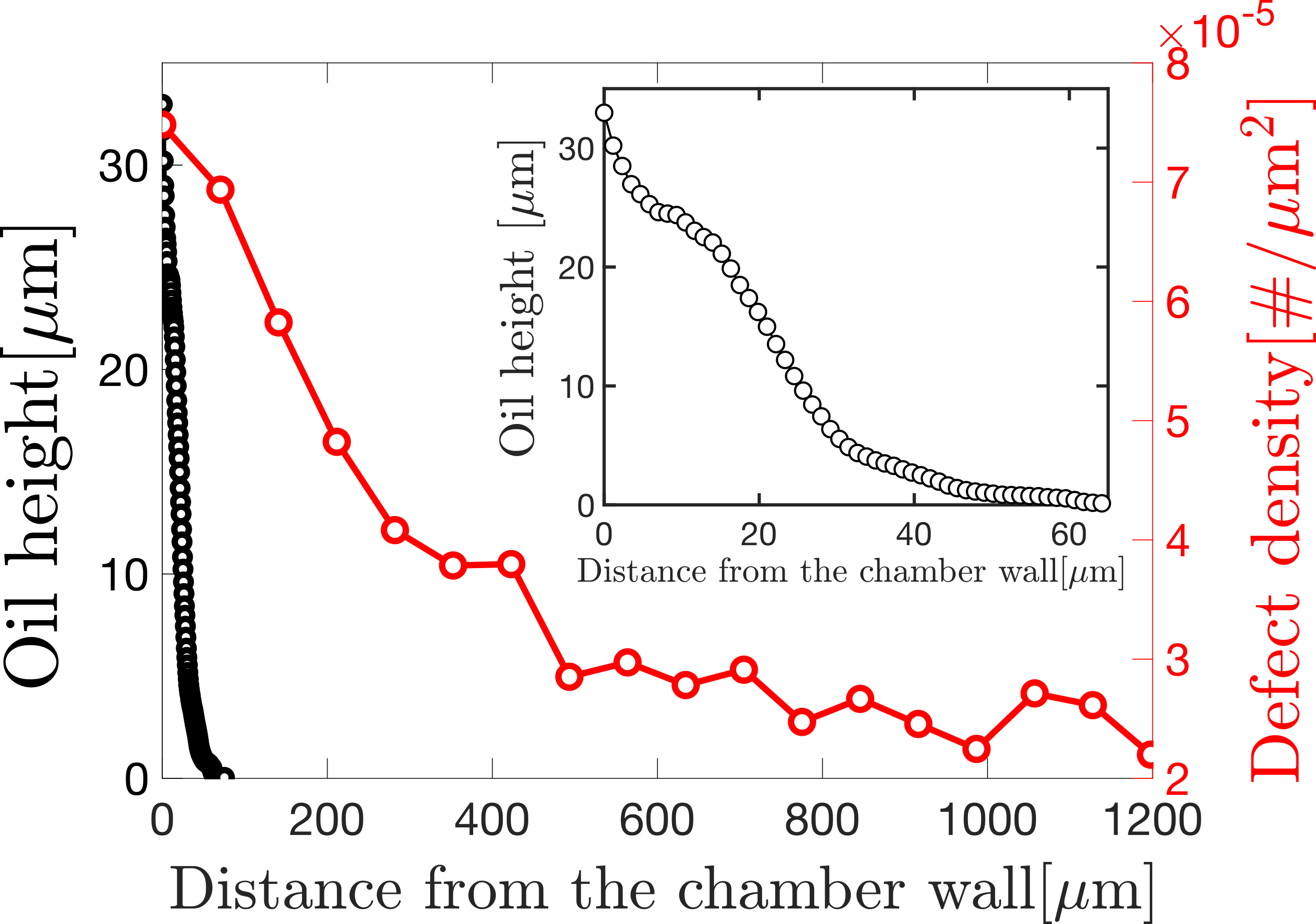}
\caption{\textbf{Relationship between the defect density and oil height with distance to the wall.} The main plot illustrates the defect density and oil height as functions of the distance to the wall, while the inset focuses on the oil height as a function of the distance to the wall. The oil height has been corrected for the oil index of refraction.} 
 \label{fig:SI_fig10}
\end{suppfigure}


\begin{suppfigure}[htbp]
\centering
 \includegraphics[width=7cm]{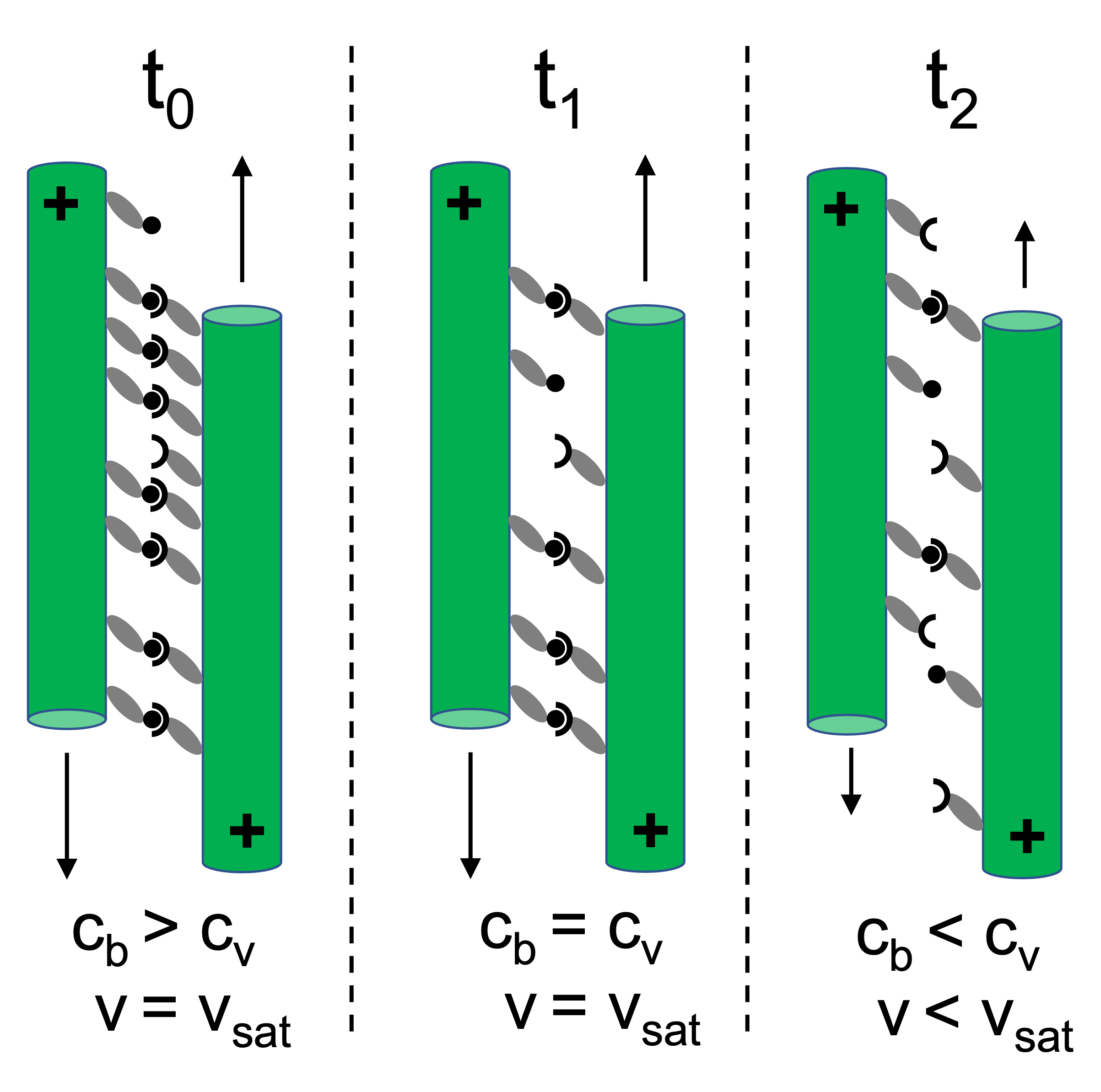}
  \caption{ Model explaining how microtubule speed and bound motor clusters change after strong illumination is extinguished. 
Initially, at $t_0$, there is a high concentration of bound motors $c_b$ and the microtubules slide at their maximum speed. After some time, $t_1$, the concentration of bound motors has decreased to $c_v$ (the concentration of bound motors at which the speed saturates), and the microtubule speed remains constant. At later times, $t_2$, the concentration of bound motors decreases below $c_v$ and the  microtubule speed decreases. }
  \label{fgr:StopRateSchematic}
\end{suppfigure}

\newpage

\setcounter{section}{7}
\section{Supplementary Videos} 

\subsection{Supplementary Video 1}
2D light activated active nematic from opto-K401 under illumination. The fluorescent images are taken with 4X 0.13 NA (Nikon Instruments, CFI Plan Fluor 4X). The exposure time was 100 msec. The scale bar is 500 $\mu$ m. The time shows min: sec, and the frame rate is 30 fps.

\begin{suppfigure}[htbp]
\centering
\includegraphics[width=6cm]{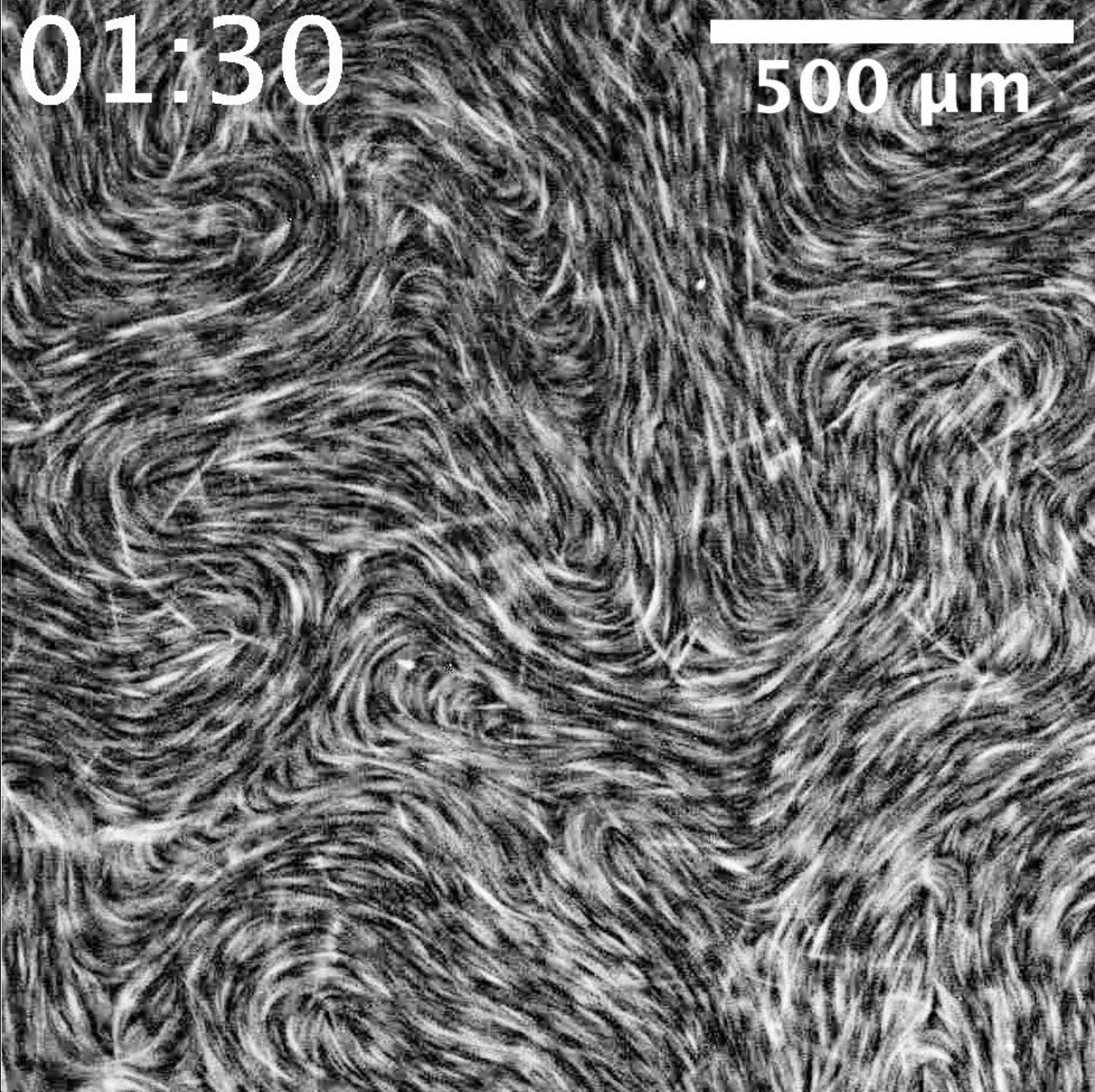}
 \label{fig:SI_fig11}
\end{suppfigure}

\subsection{ Supplementary Video 2}
2D light activated active nematic from opto-K401 under in dark. The fluorescent images are taken with 4X 0.13 NA (Nikon Instruments, CFI Plan Fluor 4X). The exposure time was 100 msec. The scale bar is 500 $\mu$ m. The time shows min: sec, and the frame rate is 30 fps.\\

\begin{suppfigure}[htbp]
\centering
\includegraphics[width=6cm]{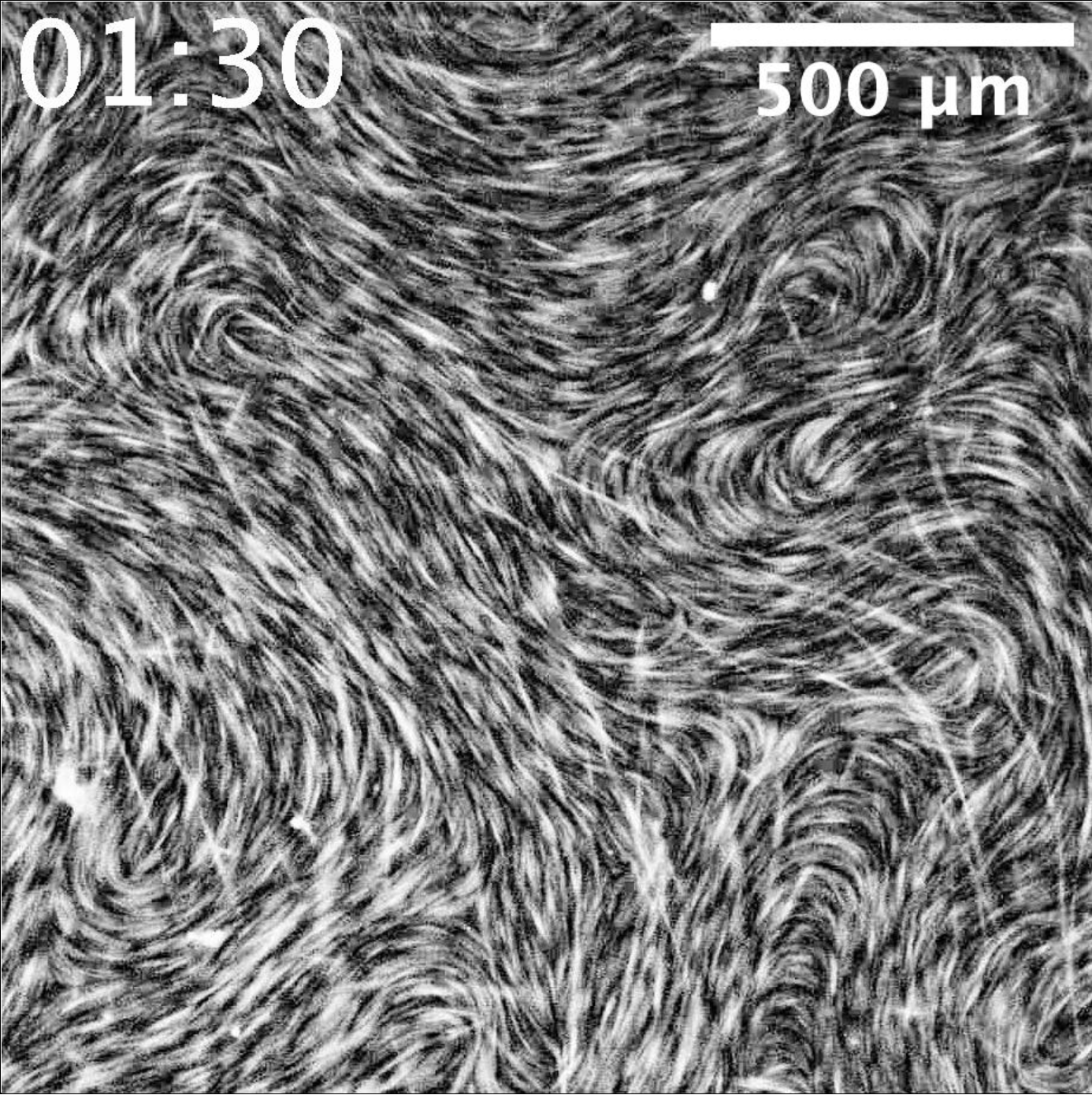}
 \label{fig:SI_fig12}
\end{suppfigure}
\newpage
\subsection{Supplementary Video 3}
Two panels. The right panel is a fluorescent image of a 2D light activated active nematic made with opto-K365 in the 3 mm by 3 mm sample geometry of Fig. 2a. The fluorescent images are taken with 4X 0.13 NA (Nikon Instruments, CFI Plan Fluor 4X). The exposure time was 200 msec. The scale bar is 500 $\mu$m. The counter represents time in Minutes: Seconds, and the frame rate is 30 fps. The left panel shows  the intensity over time. The color red represents intensity in the past and blue the intensity in the future. The red dashed line indicates the current time in the video on the right. This video shows how active nematic changes during the two-step preparation protocol shown in Fig 3. \\

\begin{suppfigure}[htbp]
\centering
\includegraphics[width=13cm]{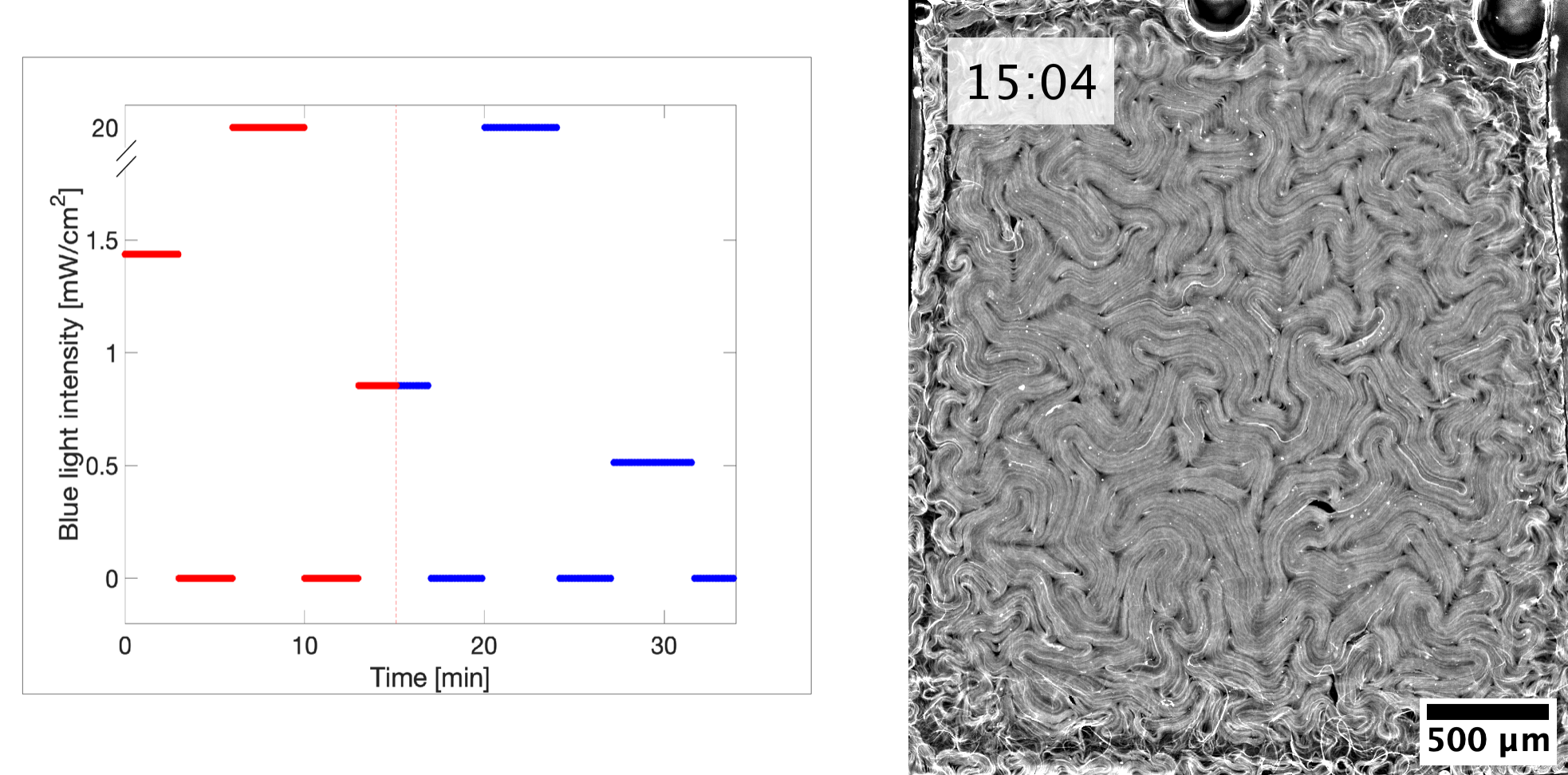}
 \label{fig:SI_fig13}
\end{suppfigure}

\end{document}